\documentclass[a4paper,11pt]{article}
\usepackage{styleBuding}
\usepackage{multirow}
\usepackage{graphicx}
\usepackage{dcolumn}
\usepackage{bm}
\usepackage{amsmath}
\usepackage{amssymb}
\usepackage{mathrsfs}
\usepackage{amsfonts}
\usepackage{color, soul}
\usepackage{makecell}
\pdfoutput=1

\let\jnfont=\rm
\def\NPB#1,{{\jnfont Nucl.\ Phys.\ B }{\bf #1},}
\def\PLB#1,{{\jnfont Phys.\ Lett.\ B }{\bf #1},}
\def\EPJC#1,{{\jnfont Eur.\ Phys.\ Jour.\ C }{\bf #1},}
\def\PRD#1,{{\jnfont Phys.\ Rev.\ D }{\bf #1},}
\def\PRL#1,{{\jnfont Phys.\ Rev.\ Lett.\ }{\bf #1},}
\def\MPLA#1,{{\jnfont Mod.\ Phys.\ Lett.\ A }{\bf #1},}
\def\JPG#1,{{\jnfont J.\ Phys.\ G}{\bf #1},}
\def\CTP#1,{{\jnfont Commun.\ Theor.\ Phys.\ }{\bf #1},}
\def\ZPC#1,{{\jnfont Z.\ Phys.\ C }{\bf #1},}
\def\JHEP#1,{{\jnfont JHEP \ }{\bf #1},}

\begin{document}

\title{\boldmath Suppressing the Scattering of WIMP DM with Nucleons in Supersymmetric Theories}


\author[a]{Junjie Cao,}
\author[a]{Lei Meng,}
\author[a]{Yuanfang Yue,}
\author[a]{Haijing Zhou,}
\author[a]{and Pengxuan Zhu}

\affiliation[a]{Department of Physics, Henan Normal University, Xinxiang 453007, China}
\emailAdd{junjiec@alumni.itp.ac.cn}
\emailAdd{mel18@foxmail.com}
\emailAdd{yuanfang405@gmail.com}
\emailAdd{zhouhaijing@htu.edu.cn}
\emailAdd{zhupx99@icloud.com}

\abstract{The continuously improving sensitivity of dark matter direct detection experiments has limited the interaction between dark matter and nucleons being increasingly feeble, while the dark matter relic density favors it to take part in weak interactions. After taking into account the constraints from the Large Hadron Collider (LHC) search for Higgs bosons and sparticles, it is becoming difficult for the neutralino dark matter in the Minimal Supersymmetric Standard Model and the Next-to-Minimal Supersymmetric Standard Model to possess these two seemingly paradoxical features in their most natural parameter space for electroweak symmetry breaking due to the limited theoretical structure. In contrast, the seesaw extension of the Next-to-Minimal Supersymmetric Standard Model, which was initially proposed to solve the neutrino mass problem, enables the lightest sneutrino to act as a viable dark matter candidate, readily has these features, and thus, it easily satisfies the constraints from dark matter and LHC experiments.
Compared with the Type-I seesaw extension, the dark matter physics in the inverse seesaw extension is more flexible, allowing it to be consistent with the experimental results in broader parameter space. We conclude that weakly interacting massive particles (such as the sneutrino in this study) work well in supersymmetric theories as dark matter candidates.
}

\maketitle

\section{Introduction\label{sec:intro}}

Current astronomical observations have confirmed the presence of dark matter (DM) and revealed that it accounts for about $27\%$ of the composition of the universe~\cite{Aghanim:2018eyx}. Among the various DM candidates, the weakly interacting massive particle (WIMP) is the most promising since it can obtain the measured DM density naturally and straightforwardly. In the popular supersymmetric theories, the lightest supersymmetric particle (LSP) usually has such a property and thus can act as a viable DM candidate~\cite{MSSM-3}. In this case, since the massive supersymmetric particles decay ultimately to the LSP due to $R$-parity conservation, their signals in colliders will contain missing momentum, a feature widely used when searching for supersymmetry at the Large Hadron Collider (LHC).

Although scientists have studied the WIMP as a popular DM candidate for many years, the rapid progress of DM direct detection (DD) experiments in recent years has sharply restricted its interaction with nucleons~\cite{Aprile:2018dbl,Aprile:2019dbj}, leading more and more people to question its rationality. Since supersymmetric theories usually predicts WIMP DMs,  they are also facing unprecedented doubts. It is the primary purpose of this work to discuss whether the WIMP DM predicted by supersymmetry can be naturally consistent with the results of the DM experiments. In particular,  although the DM density prefers it to take part in weak interactions, the continuously improved sensitivity of DM DD experiments has limited its interaction with nucleons being increasingly feeble. As shown below,  after taking into account the constraints from the LHC search for Higgs bosons and sparticles, it is becoming difficult for the neutralino DM in the Minimal Supersymmetric Standard Model (MSSM)~\cite{Gunion:1984yn,MSSM-1,MSSM-2} and the Next-to-Minimal Supersymmetric Standard Model (NMSSM)~\cite{Ellwanger:2009dp} to possess the two seemingly paradoxical features in their most natural parameter space (i.e., the Natural SUSY scenario in the literature~\cite{Baer:2012uy}) due to the limited theoretical structure. However, the seesaw extension of the NMSSM, which was initially proposed to solve the neutrino mass problem, enables the lightest sneutrino to act as a viable DM candidate and readily has these features, and thus, it easily satisfies the constraints of both dark matter and LHC experiments.

As a starting point for discussion, we first consider the MSSM~\cite{Gunion:1984yn,MSSM-1,MSSM-2}. When the correct DM density is required, the lightest neutralino with a bino field as its dominant component is customarily taken as the DM candidate\footnote{If the DM is dominated by the higgsino or wino field,  its relic density is much smaller than the measured value when $m_{DM} \lesssim 1 {\rm TeV}$ due to its relatively strong interaction with the SM particles.}.  Since the bino field transforms non-trivially under the $U(1)_Y$ group of the Standard Model (SM),  its coupling with the SM-like Higgs boson is approximated as~\cite{Pierce:2013rda,LDM-27,Cheung:2014lqa}
\begin{eqnarray}
C_{\tilde{\chi}_1^0 \tilde{\chi}_1^0 h} & \simeq & e \tan \theta_W \frac{m_Z}{\mu (1 - m_{\tilde{\chi}_1^0}^2/\mu^2)} \left ( \cos (\beta + \alpha) + \sin (\beta - \alpha) \frac{m_{\tilde{\chi}_1^0}}{\mu} \right )  \nonumber \\
& \simeq & e \tan \theta_W \frac{m_Z}{\mu (1 -  m_{\tilde{\chi}_1^0}^2/\mu^2)} \left ( \sin 2 \beta + \frac{m_{\tilde{\chi}_1^0}}{\mu} \right )
\end{eqnarray}
if the wino field is very massive, where $ m_{\tilde{\chi}_1^0}$ denotes the lightest neutralino mass that relates with the bino mass $M_1$ by $ m_{\tilde{\chi}_1^0} \simeq M_1$, $\mu$ represents higgsino mass, $\tan \beta = v_u/v_d$ is the ratio of Higgs vacuum expectation values, and
$\alpha$ is the mixing angle of the CP-even Higgs states~\cite{MSSM-2}. In obtaining the final expression, we assume the decoupling limit of the Higgs sector, i.e., $m_A \gg v $, where $m_A$ denotes the mass of the CP-odd Higgs boson. Given that the spin independent (SI) cross-section of the DM scattering with nucleons is mainly contributed to by the SM-like Higgs boson and thus approximated by~\cite{Baum:2017enm}
\begin{eqnarray}
\sigma_{\tilde{\chi}_1^0-N}^{\rm SI} \simeq 5 \times 10^{-45} {\rm cm^2} \left ( \frac{C_{\tilde{\chi}_1^0 \tilde{\chi}_1^0 h}}{0.1} \right )^2 \left (\frac{m_h}{125 {\rm GeV}} \right )^2,   \label{Approximation-SI}
\end{eqnarray}
we infer that, if $M_1$ and $\mu$ are of the same sign, $\mu$ must be sufficiently large to be consistent with the latest results of the XENON-1T experiment~\cite{Aprile:2018dbl}. Numerically speaking, we find $|\mu | \gtrsim 430~{\rm  GeV}~ (350~{\rm GeV})$ for $\tan \beta = 10~(20)$ if the DM annihilated by a resonant SM-like Higgs boson to obtain the correct relic density. These formulas also show that, if $M_1$ and $\mu$ are of opposite signs, which can result in the blind spots of the scattering~\cite{Huang:2014xua,Crivellin:2015bva,Han:2016qtc,Carena:2018nlf}, $|\mu| \sim 100~{\rm GeV}$ seems to be experimentally allowed. However, such a possibility has been limited by the XENON-1T search for spin dependent (SD) DM-nucleon scattering because, regardless of the relative sign between $M_1$ and $\mu$, small values of $|\mu|$ can enhance the scattering cross-section. This can be understood from the following approximation~\cite{Pierce:2013rda,LDM-27,Cheung:2014lqa}
\begin{eqnarray}
C_{\tilde{\chi}_1^0 \tilde{\chi}_1^0 Z} & \simeq & \frac{e \tan \theta_W \cos 2 \beta}{2} \frac{m_Z^2}{\mu^2 - m_{\tilde{\chi}_1^0}^2},  \label{MSSM-Z-coupling} \\
\sigma_{\tilde{\chi}_1^0-N}^{\rm SD} & \simeq & C_N \times \left ( \frac{C_{\tilde{\chi}_1^0 \tilde{\chi}_1^0 Z}}{0.1} \right )^2,
\end{eqnarray}
with $C_p \simeq 1.8 \times 10^{-40}~{\rm cm^2} $ for protons and $C_n \simeq 1.4 \times 10^{-40}~{\rm cm^2} $ for neutrons~\cite{Badziak:2015exr,Badziak:2017uto}. Furthermore, the LHC search for electroweakinos can play a complementary role in limiting such a possibility. In the Appendix of this work, we study these experimental constraints in the blind spot scenario of the MSSM, and find that the case of $|\mu| \lesssim 300 ~{\rm GeV}$ has been excluded for any $m_{\tilde{\chi}_1^0}$ by the experimental upper bound of the SD cross-section, and  $|\mu|$ must be larger than about $390~{\rm GeV}$ in the Higgs funnel region due to the LHC experiment. We note that a global fit of the MSSM was recently performed~\cite{Bagnaschi:2017tru}, where various experimental constraints, including those from the DM relic density, PandaX-II (2017) results for the SI cross-section~\cite{Cui:2017nnn}, PICO results for the SD cross-section~\cite{Amole:2017dex}, and the searches for supersymmetric particles at the $13~{\rm TeV}$ LHC with $36~{\rm fb^{-1}}$ data (especially the CMS analysis of the electroweakino production~\cite{Sirunyan:2018ubx}), were considered. The analysis showed that $|\mu| \gtrsim 350~{\rm GeV}$ was favored at a 95\% confidence level\footnote{We emphasize that this conclusion does not mean that the case $|\mu| < 350~{\rm GeV}$ is completely excluded by the experiments. It just implies that the case of $|\mu| < 350~{\rm GeV}$ has been tightly limited and the probability of its occurrence
is rather low in frequentist statistics. }. Such a $\mu$ can induce a tuning of about $3\%$ to predict the Z boson mass~\cite{Baer:2012uy}, and this situation will be further exacerbated if future DM DD experiments fail to detect the sign of the DM and/or forthcoming high luminosity LHC experiments do not find the evidence of the electroweakinos (see, e.g., the discussion in the Appendix about the blind spot of the MSSM).

Next we consider the NMSSM~\cite{Ellwanger:2009dp}. Since the coupling of the singlino field in this theory with SM particles may be very weak, we discuss the case of the
singlino-dominated neutralino as a DM candidate~\cite{Das:2012rr,Ellwanger:2014hia,Ellwanger:2016sur,Ellwanger:2018zxt}, which is feasible if the Higgs self-coupling coefficients $\lambda$ and $\kappa$ satisfy $\lambda > 2 \kappa$ and the gauginos are relatively massive. In the natural NMSSM that essentially requires $|\mu| \lesssim 500~{\rm GeV}$~\cite{Cao:2016nix}, this DM may annihilate by the following channels to obtain the correct relic density (see, for example, Ref.~\cite{Ellwanger:2018zxt,Ellwanger:2016sur,Cao:2013mqa,Cheung:2014lqa,Cao:2014efa,Gherghetta:2015ysa, Cao:2015loa,Badziak:2017uto,Cao:2018rix,Abdallah:2019znp,Baum:2017enm,Shang:2018dja}): $\tilde{\chi}_1^0 \tilde{\chi}_1^0 \to t \bar{t}, h_s A_s, h A_s, W W, Z Z, \cdots$ through
$s$-channel exchange of non-resonant $Z$ and Higgs bosons, $t/u$-channel exchange of electroweakinos ($h$, $h_s$, and $A_s$ denote SM-like, singlet dominated CP-even, and CP-odd Higgs bosons, respectively), the funnels in $Z$, $A_s$, $h$ and $h_s$, or the co-annihilation with sleptons, $\tilde{\chi}_2^0$ (the next-to-lightest neutralino), and $\tilde{\chi}_1^\pm$ (the lightest chargino). Among these channels, the cross-sections for the first three are potentially large, so each of them may be fully responsible for the measured DM relic density~\cite{Baum:2017enm}. However, one can verify with Effective Field Theory (EFT) that such a possibility has been tightly limited by the upper bounds of the SI and SD cross-sections (see the discussion in the Appendix).  The basic reason is that all of these channels require a large $\lambda$ to account for the density, and as a result, the DM couplings with $h$ and $Z$ bosons are sizable, since
\begin{eqnarray}
C_{\tilde{\chi}_1^0 \tilde{\chi}_1^0 h} & \simeq & \frac{\sqrt{2} m_{\tilde{\chi}_1^0} v}{\mu^2} \lambda^2 - \frac{\sqrt{2} v}{\mu} \lambda^2 \sin 2 \beta, \label{NMSSM-Higgs-Coupling} \\
C_{\tilde{\chi}_1^0 \tilde{\chi}_1^0 Z} & \simeq & \frac{e/\sin 2 \theta_W \times \left[1-\left(m_{\tilde{\chi}_1^0}/\mu\right)^2\right]
\cos 2\beta}{\left \{ 1+\left(m_{\tilde{\chi}_1^0}/\mu\right)^2-2\left(m_{\tilde{\chi}_1^0}/\mu\right)\sin2\beta
+\left[1-\left(m_{\tilde{\chi}_1^0}/\mu\right)^2\right]^2
\left({\mu}/{\lambda v}\right)^2 \right \} },  \label{NMSSM-Z-Coupling}
\end{eqnarray}
for $\lambda v /|\mu| \ll 1$ and massive gauginos~\cite{Baum:2017enm,Badziak:2015exr}, where $m_{\tilde{\chi}_1^0}$ denotes the DM mass and $v \equiv 174~{\rm GeV}$.  The other channels may play an important role in determining the density only with a specific sparticle mass spectrum, and they do not necessarily correspond to a large $\lambda$. However, given that the current sensitivities of the DM DD experiments have reached the precision of $10^{-47}~{\rm cm^2}$ for the SI cross-section~\cite{Aprile:2018dbl} and $10^{-42}~{\rm cm^2}$ for the SD cross-section~\cite{Aprile:2019dbj}, a large portion of the parameter space for the annihilations in the natural NMSSM can predict SI and/or SD cross-sections that are comparable with or even lager than their bounds. As a result, the DD experiments have tightly constrained the space~\cite{Cao:2016cnv,Cao:2018rix}. Furthermore, the LHC experiments can limit the theory, since the annihilations are usually accompanied with light sparticles or Higgs bosons~\cite{Cao:2018rix}. Recently, researchers performed comprehensive studies of the natural NMSSM by considering various experimental constraints~\cite{Cao:2018rix,Abdallah:2019znp}. In particular, they included those from the DM and LHC experiments. The conclusion was that only some corners of the parameter space are allowed, which have either of following features:
\begin{itemize}
\item $\lambda \simeq 2 \kappa$ with $\lambda \lesssim 0.05$. This case corresponds to the decoupling limit of the NMSSM~\cite{Ellwanger:2009dp}, and the DM co-annihilated with the higgsinos to obtain the density~\cite{Cao:2018rix}.
\item $\kappa \sim 0.01$, $\lambda \lesssim 0.2$ and there exists at least one light singlet dominated Higgs boson~\cite{Abdallah:2019znp}. In this case, the singlino-dominated DM annihilates in certain funnel regions, and the higgsinos decay in a complex way to satisfy the LHC constraints.
\end{itemize}
We emphasize that this conclusion is valid only for $|\mu| \lesssim 500~{\rm GeV}$, and it may be improved by more intensive study. We also emphasize that, as indicated by Eq.~(\ref{NMSSM-Higgs-Coupling}) and (\ref{NMSSM-Z-Coupling}), the increase in $|\mu|$ is helpful to relax the constraints of the DM DD experiments\footnote{Very recently we updated a previous study~\cite{Cao:2018rix}. We no longer required the fine tuning of $m_Z$ to be less than 50. Instead, we imposed the condition $\mu \leq 1~{\rm TeV}$. We found that for the singlino-dominated DM, in addition to the regions mentioned above, there is a new scenario that is consistent with current experimental constraints, which is characterized by $0.4 \lesssim \lambda \lesssim 0.7$, $200~{\rm GeV} \lesssim m_{\tilde{\chi}_1^0} \lesssim 600~{\rm GeV}$, $400~{\rm GeV} \lesssim \mu \lesssim 800~{\rm GeV}$, and the splitting between $m_{\tilde{\chi}_1^0}$ and $\mu$ is larger than approximately $80~{\rm GeV}$. This conclusion is consistent with a previous report~\cite{Baum:2017enm}. From this scenario, one can also obtain the bino-singlino well tempered DM scenario reported previously~\cite{Baum:2017enm} by choosing a negative $M_1$ that satisfies $|M_1| < \mu$, and letting $|M_1|$ approach the singlino-dominated neutralino mass from below. In either scenario, the SI cross-section can be as low as $10^{-50}~{\rm cm^2}$, while the SD cross-section is usually larger than $10^{-42}~{\rm cm^2}$. }.

Based on the discussion above, if the neutralino DM transforms non-trivially under the electroweak gauge group of the SM or it mixes with other fields and obtains the charge of the group, its scattering cross-sections with nucleons tend to be relatively large for the higgsino mass $\mu$ upper bounded by several hundred GeV. These rates usually contradict relevant experimental bounds after accounting for the rapidly improving sensitivities of the DM DD experiments in recent years. This implies that,
if one wants the scattering rate naturally suppressed, the DM should be a gauge singlet field, or its singlet component should at least be naturally far dominant over the other components. Moreover, the situation of the MSSM and NMSSM tells us that the DM is preferably not a neutralino when one extends economically the models. We emphasize that such a DM can still be a WIMP in the sense that it may have weak couplings with the particles beyond the SM, which is necessary to obtain the correct density naturally. These inferences inspire us to augment the NMSSM with a TeV-scale Type-I seesaw mechanism and choose the lightest sneutrino $\tilde{\nu}_1$ as a DM candidate~\cite{Cao:2018iyk}. The resulting theoretical framework not only produces neutrino mass, it also guarantees that $\tilde{\nu}_1$ will be almost purely right-handed, since the tiny neutrino Yukawa couplings suppress the chiral mixing of the sneutrinos significantly~\cite{Cao:2018iyk,Cerdeno:2008ep,Cerdeno:2009dv} (note that the possibility of a left-handed sneutrino as a feasible DM candidate was ruled out by DM DD experiments one decade ago~\cite{Falk:1994es,Arina:2007tm}).

Given that the right-handed sneutrino field is a gauge singlet, it can interact directly with the singlet Higgs field by triple or quartic scalar vertexes. This feature allows these fields to form a secluded DM sector, which communicates with the SM sector only by small singlet-doublet Higgs mixing and accounts for the relic density through the annihilation of $\tilde{\nu}_1$ into a pair of singlet dominated Higgs bosons~\cite{Cao:2018iyk}. In addition, the singlet Higgs field could also mediate the transition of the sneutrino pair into a higgsino pair and vice versa in the thermal bath of the early universe. If the sneutrino and the higgsino approximately degenerate in mass, the annihilation of the higgsinos in the freeze-out stage can affect the DM density significantly, which makes the relic density consistent with its experimental measurements~\cite{Cao:2018iyk} (in the literatures, this phenomenon is called co-annihilation~\cite{Griest:1990kh,Coannihilation}). Similar to the secluded DM case, $\tilde{\nu}_1$  couples very weakly with the SM particles due to its singlet nature, and this suppresses the DM scattering with nucleons. Last but not least, since $\tilde{\nu}_1$ is a scalar particle with a definite CP number, the SD cross-section of its scattering with nucleons always vanishes, which is another advantage of $\tilde{\nu}_1$ coinciding with the results of the DM DD experiments. We add that the seesaw extension of the MSSM does not have all of these features (see the introduction in~\cite{Cao:2018iyk}), and the aforementioned theoretical framework extends the field content of the NMSSM only by three generations of the right-handed neutrino superfield.  Thus, the NMSSM extension may be the most economical supersymmetric model that can suppress the DM-nucleon scattering naturally.

Similarly, one can embed the inverse seesaw mechanism in the NMSSM, and the resulting theory has similar features to the Type-I seesaw extension in DM physics~\cite{Cao:2017cjf}. Compared with the Type-I extension, this theory has at least two advantages. One is that the neutrino mass is doubly suppressed so that the right-hand neutrino mass can be naturally at the TeV scale even for sizable neutrino Yukawa couplings. The other is that DM physics is very rich due to its more complex structure in the neutrino/sneutrino sector and can be consistent with the experimental results in a more flexible way. The theory is another economic model that naturally suppresses the DM-nucleon scattering.
In fact, given that the neutralino DM has been tightly limited by DD experiments in recent years, the sneutrino DM embedded in different extensions of the MSSM
has regained broad interest~\cite{An:2011uq,BhupalDev:2012ru,Frank:2017tsm,Araz:2017qcs,Ghosh:2017yeh,Chang:2017qgi,Chatterjee:2017nyx,Ferreira:2017osm,DelleRose:2017uas,Lara:2018rwv,Zhu:2018tzm,Chang:2018agk,
Banerjee:2018uut,Ghosh:2018hly,Ghosh:2018sjz,Choi:2018vdi,Kpatcha:2019gmq,Alonso-Alvarez:2019fym,Moretti:2019yln,Faber:2019mti}.

In the NMSSM with any of the seesaw mechanisms, the scattering of the sneutrino DM with nucleons proceeds mainly by the $t$-channel exchange of CP-even Higgs particles.
Although the cross-section of the scattering is usually small, it is still potentially large if the lightest CP-even Higgs boson is significantly lighter than the SM-like Higgs boson discovered at the LHC and it contains sizable doublet Higgs components~\cite{Cao:2017cjf,Cao:2018iyk}. This case is not only consistent with current Higgs data~\cite{Cao:2019ofo}, it also has theoretical advantages of enhancing the SM-like Higgs boson mass and naturally predicting the Z boson mass~\cite{Cao:2012fz,Jeong:2014xaa}. Thus this case is attractive. Studying the characteristics of the DM-nucleon scattering in this special case for both the seesaw extensions, in particular how it coincides with the latest XENON-1T results, can improve our understanding of the scattering, which is the primary purpose of this paper. In the following, we denote the NMSSM with the Type-I seesaw mechanism as Type-I NMSSM and that with the inverse seesaw mechanism as ISS-NMSSM.

We organize this paper as follows. In Section~\ref{Section-Model}, we introduce the basics of the
sneutrino sector in the Type-I NMSSM and ISS-NMSSM, including its mass matrix, its interaction with Higgs bosons
and its scattering with nucleons. In Section~\ref{Section-results}, we consider the particular configuration of the Higgs
sector. We vary the parameters in the sneutrino sector and compare the mechanisms of the theories that keep the sneutrino
DM compatible with the DD experimental constraints. We also discuss the phenomenology of the extensions. Finally,
we draw our conclusions in Section~\ref{Section-conclusion}.

\begin{table}[t]
\begin{center}
\begin{tabular}{|c|c|c|c|c|c|}
\hline \hline
SF & Spin 0 & Spin \(\frac{1}{2}\) & Generations & \((U(1)\otimes\, \text{SU}(2)\otimes\, \text{SU}(3))\) \\
\hline
\(\hat{q}\) & \(\tilde{q}\) & \(q\) & 3 & \((\frac{1}{6},{\bf 2},{\bf 3}) \) \\
\(\hat{l}\) & \(\tilde{l}\) & \(l\) & 3 & \((-\frac{1}{2},{\bf 2},{\bf 1}) \) \\
\(\hat{H}_d\) & \(H_d\) & \(\tilde{H}_d\) & 1 & \((-\frac{1}{2},{\bf 2},{\bf 1}) \) \\
\(\hat{H}_u\) & \(H_u\) & \(\tilde{H}_u\) & 1 & \((\frac{1}{2},{\bf 2},{\bf 1}) \) \\
\(\hat{d}\) & \(\tilde{d}_R^*\) & \(d_R^*\) & 3 & \((\frac{1}{3},{\bf 1},{\bf \overline{3}}) \) \\
\(\hat{u}\) & \(\tilde{u}_R^*\) & \(u_R^*\) & 3 & \((-\frac{2}{3},{\bf 1},{\bf \overline{3}}) \) \\
\(\hat{e}\) & \(\tilde{e}_R^*\) & \(e_R^*\) & 3 & \((1,{\bf 1},{\bf 1}) \) \\
\(\hat{s}\) & \(S\) & \(\tilde{S}\) & 1 & \((0,{\bf 1},{\bf 1}) \) \\
\(\hat{\nu}\) & \(\tilde{\nu}_R^*\) & \(\nu_R^*\) & 3 & \((0,{\bf 1},{\bf 1}) \) \\
\(\widehat X\) & \(\tilde{x}\) & \(x\) & 3 & \((0,{\bf 1},{\bf 1}) \) \\
\hline \hline
\end{tabular}
\end{center}
\caption{Field content of the NMSSM with different seesaw mechanisms. The first eight fields are predicted by the NMSSM, the field $\hat{\nu}$ is
necessary for both the Type-I NMSSM and the ISS-NMSSM, and the field $\widehat X$ pertains only to the ISS-NMSSM.  }
\label{table1}
\end{table}

\section{Theoretical preliminaries}  \label{Section-Model}

As economic but complete supersymmetric theories to account for neutrino mass, the Type-I NMSSM and ISS-NMSSM adopt the same gauge groups as the NMSSM and extend each lepton generation of the NMSSM only by one and two gauge singlet chiral fields with the lepton number, respectively (see Table \ref{table1}). These fields can couple directly with the singlet Higgs field $\hat{s}$, and as a result, the singlet-dominated Higgs bosons play extraordinary roles in these theories: in addition to solving the $\mu$ problem of the MSSM~\cite{Ellwanger:2009dp} and enhancing the theoretical prediction of the SM-like Higgs boson mass through the singlet-doublet Higgs mixing~\cite{Ellwanger:2011aa, Cao:2012fz}, they are responsible for the massive neutrino mass and make the lightest sneutrino a viable DM candidate~\cite{Cao:2018iyk, Cerdeno:2008ep, Cerdeno:2009dv, Cao:2017cjf}. This feature renders the two theories quite different from seesaw extensions of the MSSM in various aspects.

The common feature of the two extensions is that they have same structure in the Higgs and neutralino/chargino sectors as that of the NMSSM, so they predict three CP-even Higgs bosons, two CP-odd Higgs bosons, five neutralinos, and two charginos. Throughout this paper, we label the particles with the same CP and spin quantum numbers in an ascending mass order, e.g., $m_{h_1} < m_{h_2} < m_{h_3}$ for the CP-even Higgs bosons. We have discussed the properties of these particles in our publications~\cite{Cao:2018iyk, Cao:2017cjf}, and here we only emphasize that they play an important role in the annihilation of the sneutrino DM, which includes the following channels~\cite{Cerdeno:2008ep, Cerdeno:2009dv}:
\begin{itemize}
\item[(1)] $\tilde{\nu}_1 \tilde{H} \rightarrow X Y$ and $\tilde{H} \tilde{H}^\prime \rightarrow X^\prime Y^\prime$, where $\tilde{H}$ and $\tilde{H}^\prime$ denote higgsino-dominated neutralinos or chargino, and $X^{(\prime)}$ and $Y^{(\prime)}$ represent any possible SM particles (including the massive neutrinos and the extra Higgs bosons if the kinematics are accessible).     This annihilation mechanism is called co-annihilation in the literature~\cite{Coannihilation, Griest:1990kh}, and it is efficient only when the mass splitting between $\tilde{H}$ and $\tilde{\nu}_1$ is less than about $10\%$. As pointed out by the Bayesian analysis of the Type-I NMSSM in~\cite{Cao:2018iyk}, it is
    the most important annihilation channel.
\item[(2)] $\tilde{\nu}_1 \tilde{\nu}_1 \rightarrow s s^\ast $, where $s$ denotes a light Higgs boson. This channel proceeds via any relevant quartic scalar coupling, the $s$-channel exchange of a Higgs boson and the $t/u$-channel exchange of a sneutrino. It is the second important annihilation channel of the DM by the analysis in~\cite{Cao:2018iyk}.
\item[(3)]  $\tilde{\nu}_1 \tilde{\nu}_1 \rightarrow V V^\ast$, $Vs $, $f \bar{f} $ where $V$ and $f$ denote any gauge boson  and SM fermion, respectively.  This kind of annihilation proceeds mainly by the $s$-channel exchange of a resonant CP-even Higgs boson.
\item[(4)]  $\tilde{\nu}_1 \tilde{\nu}_1 \rightarrow \nu_h \bar{\nu}_h$ via the $s$-channel exchange of a Higgs boson and the $t/u$-channel exchange of a neutralino, where $\nu_h$ denotes a massive neutrino.
\item[(5)] $\tilde{\nu}_1 \tilde{\nu}_1^\prime \rightarrow A_i^{(\ast)} \rightarrow X Y$ and $\tilde{\nu}_1^\prime \tilde{\nu}_1^\prime \rightarrow X^\prime Y^\prime$ where $\tilde{\nu}_1^\prime$ denotes a sneutrino with an opposite CP number to that of $\tilde{\nu}_1$.  This annihilation channel is important only when $\tilde{\nu}_1$ and $\tilde{\nu}_1^\prime$  are nearly  degenerate in mass.
\end{itemize}

The difference between the extensions and the NMSSM arises only from the neutrino/sneutrino sector, which is evident from their constructions. Since sneutrino DM is the focus of this work, we recapitulate the basics of the sneutrino sector in the following sections.

\subsection{Sneutrino sector of Type-I NMSSM}

With the field content in Table \ref{table1}, the superpotential and the soft breaking terms of the Type-I NMSSM are given by~\cite{Cerdeno:2008ep,Cerdeno:2009dv}
\begin{eqnarray}
W &=& W_{\rm MSSM} +\lambda \hat{s} \hat{H}_u \cdot  \hat{H}_d
+\frac{1}{3} \kappa \hat{s}^3+\bar{\lambda}_{\nu} \hat{s} \hat{\nu} \hat{\nu}
+ Y_\nu \,\hat{l} \cdot \hat{H}_u \,\hat{\nu},   \nonumber \\
L_{\rm soft} &=&  L_{\rm MSSM} + m_{H_d}^2 |H_d|^2 +m_{H_u}^2 |H_u|^2 +m_S^2 |S|^2 + \bar{m}_{\tilde\nu}^{2} \tilde{\nu}_{R}\tilde{\nu}^*_{R}
\nonumber \\
&& + ( \lambda A_\lambda S H_u\cdot H_d  + \frac{1}{3} \kappa A_\kappa S^3 + \bar{\lambda}_{\nu} \bar{A}_{\lambda_{\nu}}S \tilde{\nu}^*_{R} \tilde{\nu}^*_{R}
+ Y_\nu A_{\nu} \tilde{\nu}^*_{R} \tilde{l} H_u + \mbox{h.c.})
\label{superpotential}
\end{eqnarray}
where $W_{\rm MSSM}$ and $L_{\rm MSSM}$ represent the corresponding terms of the MSSM without including those for the Higgs sector,
the coefficients $\lambda$ and $\kappa$ parameterize the interactions between the Higgs fields,
$Y_\nu$ and $\bar{\lambda}_\nu$ are neutrino Yukawa couplings
with the flavor index omitted, $m_i$ ($i=H_u, H_d, \cdots$) denotes the soft breaking masses, and $A_i$ ($i=\lambda, \kappa, \cdots$) are
soft breaking parameters for the trilinear terms. Noting that the soft masses $m_{H_u}^2$, $m_{H_d}^2$, and $m_S^2$ are related to the
vacuum expectation values (vev) of the fields $H_{u}$, $H_d$ and $S$, $ \langle H_u \rangle = v_u/\sqrt{2}$, $ \langle H_d \rangle = v_d/\sqrt{2}$, and $\langle S \rangle = v_s/\sqrt{2}$, by the minimization conditions of the Higgs potential after electroweak symmetry breaking~\cite{Ellwanger:2009dp}, we
take $\lambda$, $\kappa$, $\tan \beta \equiv v_u/v_d$, $A_\lambda$, $A_\kappa$, and $\mu \equiv \lambda v_s/\sqrt{2} $ as theoretical inputs
in the following discussion.

In the Type-I NMSSM, the active neutrino mass matrix is formulated by $ m_{\nu} = \frac{1}{2}Y_{\nu} v_u M^{-1} Y_{\nu}^T v_u$, with $M = \sqrt{2} \bar{\lambda}_\nu v_s \equiv 2 \bar{\lambda}_\nu \mu/\lambda$ representing the heavy neutrino mass matrix~\cite{Kitano:1999qb}. Since the active neutrino masses are at the $0.1~{\rm eV}$ order, the magnitude of $Y_\nu$ should be around  $10^{-6}$ if the massive neutrino masses are taken at the TeV order. Moreover,
to reproduce neutrino oscillation data, $m_\nu$ must be flavor non-diagonal. This can be realized by choosing a flavor non-diagonal $Y_\nu$ and a diagonal $\bar{\lambda}_\nu$~\cite{Casas:2001sr}. If one further assumes that the soft breaking parameters in the sneutrino sector, such as $m_{\tilde{l}}$, $\bar{m}_{\tilde\nu}$, and $\bar{A}_{\lambda_{\nu}}$, are flavor diagonal, the flavor mixing in the sneutrino mass matrix is induced only by the off-diagonal elements of $Y_\nu$, which is greatly suppressed. Given these facts, it is sufficient to consider only one generation sneutrino in studying the sneutrino DM~\cite{Cao:2018iyk}.  In the following, we concentrate on the third generation case and use the symbols $\lambda_\nu$, $A_{\lambda_\nu}$, and $m_{\tilde{\nu}}$ to denote the 33 element of $\bar{\lambda}_\nu$, $\bar{A}_{\lambda_\nu}$, and $\bar{m}_{\tilde{\nu}}$, respectively.

After rephrasing the sneutrino fields by CP-even and CP-odd parts,
\begin{equation}
\tilde{\nu}_L \equiv \frac{1}{\sqrt{2}}(\tilde{\nu}_{L1} + i
\tilde{\nu}_{L2}) ,
\quad\quad
\tilde{\nu}_R \equiv \frac{1}{\sqrt{2}}(\tilde{\nu}_{R1} + i \tilde{\nu}_{R2}) ,
\end{equation}
the sneutrino mass matrix in the bases ($\tilde{\nu}_{L1}$, $\tilde{\nu}_{R1}$, $\tilde{\nu}_{L2}$, $\tilde{\nu}_{R2}$) is given by
\begin{eqnarray}
{\cal M}^2_{\tilde{\nu}} =\begin{pmatrix}
m_{L\bar{L}}^2
& \frac{m_{LR}^2+m_{L\bar{R}}^2 + {\rm c.c} }{2}
&  0
& i \frac{m_{LR}^2-m_{L\bar{R}}^2 - {\rm c.c} }{2}  \\
\frac{m_{LR}^2+m_{L\bar{R}}^2 + {\rm c.c} }{2}
& m_{R\bar{R}}^2 + m_{RR}^2+m_{RR}^{2*}
& i \frac{m_{LR}^2-m_{L\bar{R}}^2 - {\rm c.c} }{2}
& i (m_{RR}^2 - m_{RR}^{2*})  \\
0
& i \frac{m_{LR}^2-m_{L\bar{R}}^2 - {\rm c.c} }{2}
& m_{L\bar{L}}^2
& \frac{-m_{LR}^2+m_{L\bar{R}}^2 + {\rm c.c} }{2} \\
i \frac{m_{LR}^2-m_{L\bar{R}}^2 - {\rm c.c} }{2}
& i (m_{RR}^2 - m_{RR}^{2*})
& \frac{-m_{LR}^2+m_{L\bar{R}}^2 + {\rm c.c} }{2}
& m_{R\bar{R}}^2 - m_{RR}^2 - m_{RR}^{2*}
\end{pmatrix},
\label{sneutrino_matrix}
\end{eqnarray}
where
\begin{eqnarray}
  m^2_{L\bar{L}}&\equiv& m^2_{\tilde{l}}+\frac{1}{2}|Y_{\nu}v_{u}|^2+\frac{1}{8}(g^2_1+g^2_2)(v^2_d-v^2_u) , \nonumber \\
  m^2_{LR}&\equiv& Y_{\nu}v_u(\lambda_{\nu}v_s)^* , \nonumber \\
  m^2_{L\bar{R}}&\equiv& \frac{1}{2}Y_{\nu}[(-\lambda v_sv_d)^*+\sqrt{2}A_{Y_{\nu}}v_u] , \nonumber \\
  m^2_{R\bar{R}}&\equiv& m^2_{\tilde{\nu}} + 2|\lambda_{\nu}v_s|^2+\frac{1}{2}|Y_{\nu}v_u|^2 , \nonumber \\
  m^2_{RR}&\equiv& \frac{1}{2}\lambda_{\nu}[\sqrt{2}A_{\lambda_{\nu}}v_s+(\kappa v^2_s-\lambda v_dv_u)^*].
\label{sn:mrr}
\end{eqnarray}
If all the parameters in the matrix are real, i.e., there is no CP violation, the real and imaginary parts of the sneutrino fields will not mix. In this case, the $4 \times 4$ mass matrix splits into two $2 \times 2$ matrices:
\begin{eqnarray}
&&\frac{1}{2}(\tilde{\nu}_{Li}, \tilde{\nu}_{Ri})
\left(
\begin{array}{cc}
m_{L\bar{L}}^2           &  \pm m_{LR}^2+m_{L\bar{R}}^2 \\
\pm m_{LR}^2+m_{L\bar{R}}^2  &  m_{R\bar{R}}^2 \pm 2m_{RR}^2 \\
\end{array}
\right)
\left(
\begin{array}{c}
\tilde{\nu}_{Li}  \\
\tilde{\nu}_{Ri} \\
\end{array}
\right), \nonumber
\end{eqnarray}
where $i=1$ and $2$ denote CP-even and CP-odd states, respectively, and the minus signs in the matrix elements are relevant to the CP-odd states.
These formulas indicate that the chiral mixings of the sneutrinos are proportional to $Y_{\nu}$, and hence, they are negligible. Thus,
the sneutrino mass eigenstate and the chiral state coincide. In our study, we selected the lightest right-handed state as the DM candidate.
The formulas also indicate that the mass splitting between CP-even and CP-odd right-handed states  is given by $\Delta m^2 \equiv m_{\rm even}^2 - m_{\rm odd}^2 = 4 m^2_{RR}$, which
implies that the sneutrino DM has a definite CP number, i.e. it is CP-even if $m^2_{RR} < 0$ and CP-odd for the other case. In our study of the sneutrino DM, we consider both CP possibilities.

The coupling strengths of the CP-even sneutrino DM with Higgs bosons are given by
\begin{eqnarray}
C_{\tilde{\nu}_1\tilde{\nu}_1 h_i}&=&
\frac{\lambda\lambda_{\nu}M_W}{g}({\rm sin\beta} Z_{i1} + {\rm cos\beta} Z_{i2}) - \left[
\frac{\sqrt{2}}{\lambda} \left(2\lambda_{\nu}^2 + \kappa\lambda_{\nu}\right) \mu  - \frac{\lambda_{\nu} A_{\lambda_\nu}}{\sqrt{2}}
\right]Z_{i3}, \label{Csnn} \\
C_{\tilde{\nu}_1 \tilde{\nu}_1 h_i h_j} &=& \frac{1}{2} \lambda \lambda_\nu Z_{i1} Z_{j2} - (\lambda_\nu^2 + \frac{1}{2} \lambda_\nu \kappa )  Z_{i3} Z_{j3}, \nonumber \\
C_{\tilde{\nu}_1 \tilde{\nu}_1 A_m A_n} &=& -  \frac{1}{2} \lambda \lambda_\nu \cos \beta \sin \beta Z^\prime_{m1} Z^\prime_{n1} - (\lambda_\nu^2 - \frac{1}{2} \lambda_\nu \kappa )  Z^\prime_{m2} Z^\prime_{n2}, \nonumber
\end{eqnarray}
where $Z_{ij}$ ($i,j=1,2,3$) and $Z^\prime_{mn}$ ($m,n=1,2$) are the elements of the rotations to diagonalize the CP-even Higgs mass matrix in the bases (${\rm Re}[H_d^0]$, ${\rm Re}[H_u^0]$, ${\rm Re}[S]$) and the CP-odd Higgs mass matrix in the bases ($A \equiv \cos \beta {\rm Im}[H_u^0] - \sin \beta {\rm Im}[H_d^0]$, ${\rm Im}[S]$), respectively.
Those for the CP-odd DM case are obtained from the formulas for the CP-even state by the substitution $\lambda_\nu \to -\lambda_\nu$. These expressions indicate that $C_{\tilde{\nu}_1\tilde{\nu}_1 h_i}$ is suppressed by a factor $\lambda \lambda_\nu \cos \beta$ if $h_i$ is the SM Higgs boson, which corresponds to setting $Z_{i1} = \cos \beta$, $Z_{i2} = \sin \beta$,  and $Z_{i3} = 0$, and all three couplings may be moderately large only if the Higgs bosons are singlet dominant. Moreover, among the parameters in the sneutrino sector, $\lambda_\nu$ and $A_{\lambda_\nu}$ affect both the couplings and masses of the sneutrinos,
while $m_{\tilde{\nu}}^2$ only affects the masses. These features are helpful for understanding the behavior of the DM-nucleon scattering discussed below.

\subsection{Sneutrino sector of ISS-NMSSM}

Compared with the Type-I NMSSM, the ISS-NMSSM is much more complex in its neutrino/sneutrino sector. With the assignment of the quantum number for the fields in Table \ref{table1}, its renormalizable superpotential and soft breaking terms take the following form~\cite{Cao:2017cjf}
\begin{eqnarray}
 W &= & \left [ W_{\rm MSSM}+\lambda\,\hat{s}\,\hat{H}_u \cdot \, \hat{H}_d\,
 +\frac{1}{3} \kappa \,\hat{s}^3 \right ] +  \left [\frac{1}{2} \mu_X \,\widehat X\,\widehat X\,+  \lambda_\nu \,\hat{s}\,\hat{\nu}_R \,\widehat X\,
 +Y_\nu \,\hat{l} \cdot \hat{H}_u \,\hat{\nu}_R \right ], \nonumber  \\
L_{\rm soft} &=& \left [ L_{\rm MSSM} + m_S^2 |S|^2 +  \lambda A_{\lambda} S H_u\cdot H_d + \frac{\kappa}{3} A_{\kappa} S^3 \right ] \nonumber  \\
&& + \left [ m_{\tilde{\nu}}^{2} \tilde{\nu}_{R}\tilde{\nu}^*_{R} +  m_{\tilde{x}}^{2} \tilde{x}\tilde{x}^* + \frac{1}{2} B_{\mu_X} \tilde{x} \tilde{x} +  (\lambda_{\nu} A_{\lambda_\nu} S \tilde{\nu}^*_{R} \tilde{x} + Y_\nu A_{\nu} \tilde{\nu}^*_{R} \tilde{l} H_u + \mbox{h.c.}) \right ], \nonumber
\end{eqnarray}
where the terms in the first bracket on the right side make up the Lagrangian of the NMSSM, and those in the second
bracket are needed to implement the inverse seesaw mechanism. The coefficients $\lambda_\nu$ and $Y_\nu$ in the superpotential are neutrino Yukawa couplings,
$A_{\lambda_\nu}$ and $A_{\nu}$ in $L_{soft}$ are coefficients for the soft breaking trilinear terms, and all of them are
$3 \times 3$ (diagonal or non-diagonal) matrices in the flavor space. Moreover, among the parameters in the superpotential,
only the matrix  in the flavor space $\mu_X$ is dimensional. This matrix parameterizes the effect of the lepton number violation (LNV), which may
arise from the integration of heavy particles in an ultraviolet high-energy theory with LNV interactions
(see, for example, \cite{ISS-NMSSM-1,ISS-NMSSM-3,ISS-NMSSM-2}), so the magnitude of its elements should be suppressed.
Similarly, the soft breaking parameter $B_{\mu_X}$ tends to be small.

By defining $M_D = \frac{v_u}{\sqrt{2}} Y_\nu $, $M_R = \frac{v_s}{\sqrt{2}} \lambda_\nu $, and $\|M\| \equiv \sqrt{{\rm Tr}(M^\dag M)}$ for an arbitrary matrix $M$,
one can approximate the mass matrix of the light active neutrinos by
\begin{eqnarray}
M_\nu \simeq \left[M_D^T M_R^{T^{-1}}\right]\mu_X \left[(M_R^{-1})M_D\right] \equiv F \mu_X F^T \,  \label{active-neutrino-mass}
\end{eqnarray}
under the condition $\|\mu_X\| \ll \|M_D\| \ll \|M_R\|$. In this approximation, $F \equiv M_D^T M_R^{T^{-1}}$, and the magnitudes of its elements are of the order $\|M_D\|/\|M_R\|$.
Thus, in the inverse seesaw mechanism, the active neutrino masses are suppressed in a double way, i.e., by the smallness of
the LNV matrix $\mu_X$ and also by the suppression factor $\|M_D\|^2/\|M_R\|^2$.
For $\|\mu_X \| \sim {\cal O}({\rm KeV})$, one can easily conclude that $ \| Y_\nu \| \sim {\cal O}(0.1)$
if the mass scale of the massive neutrinos $\|M_R \|$ is taken at the TeV order.

Given the expression in Eq.~(\ref{active-neutrino-mass}), one can solve the matrix $\mu_X$ with  neutrino masses $m_{\nu_i}$
and the unitary Pontecorvo-Maki-Nakagawa-Sakata matrix $U_{\rm PMNS}$ extracted from low energy experiments~\cite{Tanabashi:2018oca}
to obtain~\cite{Arganda:2014dta,Baglio:2016bop}:
\begin{eqnarray}
\mu_X=M_R^T ~m_D^{T^{-1}}~ U_{\rm PMNS}^* \, m_\nu^{\rm diag} \,  U_{\rm PMNS}^\dagger~ {m_D}^{-1} M_R,  \nonumber
\end{eqnarray}
with $m_\nu^{\rm diag} = \mathrm{diag}(m_{\nu_1}\,, m_{\nu_2}\,, m_{\nu_3})$.
This formula indicates that if the Yukawa couplings $Y_\nu$ and $\lambda_\nu$ are flavor diagonal,
the neutrino oscillation observed in the low energy experiments is attributed only to the non-diagonality of $\mu_X$.
In this case, being unitary in the neutrino sector requires~\cite{Cao:2019aam}
\begin{eqnarray}
\frac{[\lambda_{\nu}]_{11} \mu}{{[Y_{\nu}]_{11} \lambda v_u}} > 14.1, \quad \frac{[\lambda_{\nu}]_{22} \mu}{{[Y_{\nu}]_{22} \lambda v_u}} > 33.7, \quad \frac{[\lambda_{\nu}]_{33} \mu}{{[Y_{\nu}]_{33} \lambda v_u}}  > 9.4,  \label{unitaryconstriants}
\end{eqnarray}
which reveals that the ratio $[\lambda_{\nu}]_{33}/[Y_{\nu}]_{33}$ may be significantly smaller than $[\lambda_{\nu}]_{11}/[Y_{\nu}]_{11}$ and $[\lambda_{\nu}]_{22}/[Y_{\nu}]_{22}$ once $\lambda$, $\mu$, and $v_u$ (or alternatively $\tan \beta$) are given. Furthermore, if $\lambda_\nu$ is assumed proportional to the identity matrix, $[Y_\nu]_{33}$ may be much larger than $[Y_\nu]_{11}$ and $[Y_\nu]_{22}$.

In addition to the inputs $\lambda$, $\kappa$, $\mu$, and $\tan \beta$ in the Higgs sector, the sneutrino sector in the ISS-NMSSM involves the parameters $Y_\nu$, $\lambda_\nu$, $A_\nu$, $A_{\lambda_\nu}$, $\mu_X$, and $B_{\mu_X}$ and the soft breaking masses $m_{\tilde{l}}$, $m_{\tilde{\nu}}$, and $m_{\tilde{x}}$. As a result, the squared mass of the sneutrino fields is given by a $9 \times 9$ matrix in three-generation $(\tilde{\nu}_L, \tilde{\nu}_R^\ast, \tilde{x})$ bases, whose form is quite complicated.  However, we note the fact that among these parameters, only $\mu_X$  must be flavor non-diagonal to predict the neutrino oscillations, but since its elements are usually less than 10~KeV~\cite{Arganda:2014dta}, it can be safely neglected in calculating the sneutrino mass. Thus, if there are no flavor mixings for the other parameters, the matrix is flavor diagonal, and one generation $(\tilde{\nu}_L, \tilde{\nu}_R^\ast, \tilde{x})$ bases can be used to study the properties of the sneutrino DM. In this work, we take the third generation sneutrinos as the DM sector, which is motivated by the fact that both the unitary bound and the constraints of the LHC search for sparticles in this sector are significantly weaker than those in the other generations~\cite{Cao:2017cjf}. In the discussion below, when we refer to the parameters $Y_\nu$, $\lambda_\nu$, $A_{\nu}$, $A_{\lambda_\nu}$, $m_{\tilde{\nu}}$, $m_{\tilde{x}}$, and $m_{\tilde{l}}$, we are actually referring their 33 elements, which is the same as what we did for the Type-I NMSSM.

If the sneutrino field are decomposed into CP-even and CP-odd parts,
\begin{eqnarray}
\tilde{\nu}_{L} =  \, \frac{1}{\sqrt{2}} \left ( \phi_1  + i \sigma_1 \right ),~~~~
\tilde{\nu}_{R} = \frac{1}{\sqrt{2}} \left (\phi_{2}  + i \sigma_{2} \right ), ~~~~ \tilde{x} = \frac{1}{\sqrt{2}}
\left ( \phi_{3}  + i \sigma_{3} \right ),
\end{eqnarray}
the squared mass matrix of the CP-even fields is given  by
\begin{eqnarray}
m^2_{\tilde{\nu}} = \left(
\begin{array}{ccc}
m_{11} \quad &m_{12} \quad & m_{13} \\
m_{12}^\ast \quad  &m_{22} \quad &m_{23}\\
m_{13}^\ast \quad &m_{23}^\ast \quad &m_{33} \end{array}
\right),
\label{sneutrino-mass}
 \end{eqnarray}
in the bases $(\phi_1, \phi_2, \phi_3)$, where
\begin{eqnarray}
m_{11} &=& \frac{1}{4} \left [ 2 v_{u}^{2} {{\rm Re} \Big({Y_{\nu}  Y_\nu^*}\Big)}  + 4 {{\rm Re} \Big(m_{\tilde{l}}^2\Big)} \right ]  + \frac{1}{8} \Big(g_{1}^{2} +
g_{2}^{2}\Big) \Big(- v_{u}^{2}  + v_{d}^{2}\Big) {\bf 1},  \nonumber \\
m_{12} &=& -\frac{1}{2} v_d v_s {{\rm Re} \Big(\lambda Y_\nu^\ast \Big)}  + \frac{1}{\sqrt{2}} v_u {{\rm Re}\Big(Y_\nu A_\nu \Big)}, \nonumber  \\
m_{13} &=& \frac{1}{2} v_s v_u {{\rm Re}\Big({Y_{\nu}  \lambda_\nu^\ast}\Big)}, \nonumber \\
m_{22} &=& \frac{1}{4} \left [ 2 v_{s}^{2} {{\rm Re}\Big({\lambda_\nu  \lambda_{\nu}^{\ast}}\Big)}  + 2 v_{u}^{2} {{\rm Re}\Big({Y_\nu  Y_{\nu}^{\ast}}\Big)}
+ 4 {{\rm Re}\Big(m_{\tilde{\nu}}^2\Big)} \right ],  \nonumber \\
m_{23} &=& \frac{1}{8} \left \{ -2 v_d v_u \lambda \lambda_{\nu}  + 2 \left [ \Big(- v_d v_u \lambda  + v_{s}^{2} \kappa \Big)\lambda_{\nu}^{\ast}  + v_{s}^{2} \kappa \lambda_{\nu} \right ] \right . \nonumber \\
&& \quad \quad \quad \quad \quad \quad \quad \quad \quad \quad \quad \quad \left .  + \sqrt{2} v_s \left [ -4 {{\rm Re}\Big({\mu_X  \lambda_{\nu}^{\ast}}\Big)}
+ 4 {{\rm Re}\Big(A_{\lambda_\nu}^{\ast} \lambda_\nu \Big)} \right ] \right \}, \nonumber \\
m_{33} &=& \frac{1}{8} \Big(4 v_{s}^{2} {{\rm Re}\Big({\lambda_{\nu}  \lambda_\nu^\ast}\Big)} + 8 {{\rm Re}\Big(B_{\mu_X}\Big)}  + 8 {{\rm Re}\Big({\mu_X  \mu_X^\ast}\Big)}  + 8 {{\rm Re}\Big(m_{\tilde{x}}^2\Big)} \Big).
\end{eqnarray}
This matrix shows that the mixing of the $\phi_1$ field with the other fields is determined by the parameters $Y_{\nu}$ and $A_\nu$.
As $Y_{\nu}$ approaches zero, $m_{12}$ and $m_{13}$ vanish, and consequently, $\phi_1$ no longer mixes
with the fields $\phi_2$ and $\phi_3$. This situation is quite similar to that of the Type-I NMSSM. Moreover, if the first terms in $m_{22}$ and $m_{33}$ are far dominant over the other terms in their respective expressions, then $m_{22} \simeq m_{33}$. This results in maximal mixing between the $\phi_2$ and $\phi_3$ fields. In this case, $\tilde{\nu}_1$ is approximated by
$\tilde{\nu}_1 \simeq 1/\sqrt{2} [ \phi_2 - {\rm sgn}(m_{23}) \phi_3 ] $ if the left-handed field is decoupled~\cite{Cao:2017cjf}. Such a situation is frequently encountered in the ISS-NMSSM.

The mass matrix of the CP-odd sneutrino fields can be obtained from that of the CP-even fields by the substitution $\mu_X \to -\mu_X$ and $B_{\mu_X} \to - B_{\mu_X}$.
Since $\mu_X$ and $B_{\mu_X}$ represent the degree of the LNV, their effect on $m_{33}$ should be much smaller than the other contributions. In the extreme case of $\mu_X = 0$ and $B_{\mu_X} = 0$, any CP-odd sneutrino state is accompanied by a mass-degenerate CP-even sneutrino state. Consequently, the sneutrino particle as a mass eigenstate corresponds to a complex field, and it has an anti-particle~\cite{MSSM-ISS-1}. Alternatively, if $B_{\mu_X}$ takes a naturally suppressed value, the mass splitting between the CP-even and CP-odd states is usually tiny, e.g., less than $0.2~{\rm GeV}$ for $B_{\mu_X} = 100~{\rm GeV^2}$ and $m_{\tilde{\nu}_1} \sim 100~{\rm GeV}$. Such a sneutrino is called a pseudo-complex particle in the literature~\cite{MSSM-ISS-2,ISS-NMSSM-3,MSSM-ISS-10,MSSM-ISS-15}. This feature of the sneutrino DM is quite different from that in the Type-I NMSSM.

In the ISS-NMSSM, the $\tilde{\nu}_1^\ast \tilde{\nu}_1 h_i$ coupling strength is given by
\begin{eqnarray}
C_{\tilde{\nu}_1^\ast \tilde{\nu}_1 h_i} =  C_{\tilde{\nu}_1^\ast \tilde{\nu}_1 H_d} Z_{i1}  +  C_{\tilde{\nu}_1^\ast \tilde{\nu}_1 H_u} Z_{i2} +  C_{\tilde{\nu}_1^\ast \tilde{\nu}_1 S} Z_{i3},  \nonumber
\end{eqnarray}
where $C_{\tilde{\nu}_1^\ast \tilde{\nu}_1 s}$ ($s={\rm Re}[H_d^0], {\rm Re}[H_u^0], {\rm Re}[S]$) denotes the coupling of $\tilde{\nu}_1$ with the scalar field $s$. For one generation sneutrino case, it is given by
\begin{eqnarray}
C_{\tilde{\nu}_1^\ast \tilde{\nu}_1 {\rm Re}[H_d^0]} &=& \lambda Y_\nu v_s V_{11} V_{12} + \lambda \lambda_\nu v_u V_{12} V_{13} - \frac{1}{4} (g_1^2 + g_2^2) v_d V_{11} V_{11}, \nonumber \\
C_{\tilde{\nu}_1^\ast \tilde{\nu}_1 {\rm Re}[H_u^0]} &=&  \lambda \lambda_\nu v_d V_{12} V_{13} - \sqrt{2} Y_\nu A_\nu V_{11} V_{12} - Y_\nu^2 v_u  V_{11} V_{11} - \lambda_\nu Y_\nu v_s V_{11} V_{13} \nonumber \\ && \quad \quad - Y_\nu^2 v_u V_{12} V_{12} + \frac{1}{4} (g_1^2 + g_2^2) v_u V_{11} V_{11},  \nonumber \\
C_{\tilde{\nu}_1^\ast \tilde{\nu}_1 {\rm Re}[S]} &=& \lambda Y_\nu v_d V_{11} V_{12} - 2 \kappa \lambda_\nu v_s  V_{12} V_{13} - \sqrt{2} \lambda_\nu A_{\lambda_\nu} V_{12} V_{13}  + \sqrt{2} \lambda_\nu \mu_X V_{12} V_{13} \nonumber \\ && \quad \quad -  \lambda_\nu Y_\nu v_u V_{11} V_{13} - \lambda_\nu^2 v_s (V_{12} V_{12} + V_{13} V_{13} ),   \label{Coupling-expression}
\end{eqnarray}
where $V$ denotes the rotation matrix to diagonalize the squared mass matrix in Eq.~(\ref{sneutrino-mass}).  This formula indicates that, among the parameters in the sneutrino sector, $Y_\nu$, $\lambda_\nu$, $A_\nu$, and $A_{\lambda_\nu}$ affect not only the interactions of the sneutrinos but also the mass spectrum and the mixing of the sneutrinos. In contrast, the soft breaking masses $m_{\tilde{\nu}}^2$ and $m_{\tilde{x}}^2$ affect only the latter property. Given the typical value of the quantities in Eq.~(\ref{Coupling-expression}), i.e., $\tan \beta \gg 1$, $|V_{11}| < 0.1$, $Y_\nu, \kappa, \lambda, \lambda_\nu \sim {\cal{O}}(0.1)$ and $\lambda_\nu v_s, \lambda v_s, A_\nu, A_{\lambda_\nu} \sim {\cal{O}}({\rm 100~GeV})$, the couplings $C_{\tilde{\nu}_1 \tilde{\nu}_1 S}$ can be approximated by
\begin{eqnarray}
C_{\tilde{\nu}_1^\ast \tilde{\nu}_1 {\rm Re}[H_d^0]} &\simeq & \lambda Y_\nu v_s V_{11} V_{12} + \lambda \lambda_\nu v_u V_{12} V_{13}, \nonumber \\
C_{\tilde{\nu}_1^\ast \tilde{\nu}_1 {\rm Re}[H_u^0]} &\simeq &  - \sqrt{2} \lambda_\nu A_\nu V_{11} V_{12} - \lambda_\nu Y_\nu v_s V_{11} V_{13} - Y_\nu^2 v_u V_{12} V_{12},  \nonumber \\
C_{\tilde{\nu}_1^\ast \tilde{\nu}_1 {\rm Re}[S]} &\simeq & - 2 \kappa \lambda_\nu v_s  V_{12} V_{13} - \sqrt{2} \lambda_\nu A_{\lambda_\nu} V_{12} V_{13} - \lambda_\nu^2 v_s,  \label{approximation-1}
\end{eqnarray}
and it can be estimated that $|C_{\tilde{\nu}_1^\ast \tilde{\nu}_1 {\rm Re}[H_d^0]}|, |C_{\tilde{\nu}_1^\ast \tilde{\nu}_1 {\rm Re}[H_u^0]}| \lesssim 10~{\rm GeV}$ and $C_{\tilde{\nu}_1^\ast \tilde{\nu}_1 {\rm Re}[S]} \lesssim 100~{\rm GeV}$. This estimation reflects the fact that $|C_{\tilde{\nu}_1^\ast \tilde{\nu}_1 {\rm Re}[S]}|$ is usually much larger than the other two couplings. The basic reason for this is that $\tilde{\nu}_1$ is a singlet-dominated scalar, so it can couple directly with the field $S$, while in the case of $V_{11} = 0$, the other couplings emerge only after electro-weak symmetry breaking.

\subsection{DM-nucleon scattering}

Since the sneutrino DM in the NMSSM extensions is a singlet-dominated scalar with definite CP and lepton numbers, its interaction with nucleon $N$ ($N=p,n$) is mediated mainly by the CP-even Higgs bosons $h_i$ ($i=1,2,3$), yielding the effective operator ${\cal{L}}_{\tilde{\nu}_1 N} = f_N \tilde{\nu}_1^\ast \tilde{\nu}_1 \bar{\psi}_N \psi_N$, where the coefficient $f_N$ is given by~\cite{Han:1997wn}
\begin{eqnarray*}
f_N &=&  m_N \sum_{i=1}^3 \frac{C_{\tilde{\nu}_1^\ast \tilde{\nu}_1 h_i}}{m_{h_i}^2} C_{h_i N N} =  m_N \sum_{i=1}^3 \frac{C_{\tilde{\nu}_1^\ast \tilde{\nu}_1 h_i}}{m_{h_i}^2} \frac{(-g)}{2 m_W} \left ( \frac{Z_{i2}}{\sin\beta} F^{N}_u +  \frac{Z_{i1}}{\cos \beta} F^{N}_d \right ),
\end{eqnarray*}
and $C_{h_i N N}$ denotes the Yukawa coupling of the Higgs boson $h_i$ with the nucleon $N$ that relies on the nucleon form factors $f^{N}_G=1-\sum_{q=u,d,s}f^{N}_q$, $F^{N}_u=f^{N}_u+\frac{4}{27}f^{N}_G$
and $F^{N}_d=f^{N}_d+f^{N}_s+\frac{2}{27}f^{N}_G$ with $f^{N}_q=m_N^{-1}\left<N|m_qq\bar{q}|N\right>$
(for $q=u,d,s$). This operator does not contribute to the SD cross-section for the  $\tilde{\nu}_1 - N$ scattering, while it predicts the SI cross-section as follows~\cite{Han:1997wn}:
\begin{eqnarray}
\sigma^{\rm SI}_{\tilde{\nu}_1-N} &=& \frac{\mu^2_{\rm red}}{4 \pi m_{\tilde{\nu}_1}^2} f_N^2 = \frac{4 F^{N2}_u \mu^2_{\rm red} m_N^2}{\pi}
\left \{ \sum_i ({a_{u}}_i  + {a_{d}}_i F^{N}_d/F^{N}_u  ) \right \}^2,
 \label{SI-expression1}
\end{eqnarray}
where $\mu_{\rm red}= m_N/( 1+ m_N/m_{\tilde{\nu}_1}) $ is the reduced mass of the nucleon with $m_{\tilde{\nu}_1}$, and the quantities
${a_{u}}_i$ and ${a_{d}}_i$ are defined by
\begin{eqnarray}
{a_{u}}_i = -\frac{g}{8 m_W} \frac{C_{\tilde{\nu}_1^\ast \tilde{\nu}_1 h_i}}{m_{h_i}^2 m_{\tilde{\nu}_1}} \frac{Z_{i2}}{\sin\beta}, \quad \quad
{a_{d}}_i = -\frac{g}{8 m_W} \frac{C_{\tilde{\nu}_1^\ast \tilde{\nu}_1 h_i}}{m_{h_i}^2 m_{\tilde{\nu}_1}} \frac{Z_{i1}}{\cos \beta},  \label{scattering-form}
\end{eqnarray}
to facilitate our analysis\footnote{In the case that the DM candidate is a Majorana fermion, e.g., the lightest neutralino in the MSSM and NMSSM, the scattering cross-section takes the same form as Eq.~(\ref{SI-expression1}) except that $a_{q_i}$ is obtained from Eq.~(\ref{scattering-form}) by the replacement $C_{\tilde{\nu}_1^\ast \tilde{\nu}_1 h_i}/m_{\tilde{\nu}_1} \to C_{\tilde{\chi}_1^0 \tilde{\chi}_1^0 h_i}$~\cite{Cao:2017cjf}. This similarity was used to compare the scattering rate in different models~\cite{Cao:2017cjf}. }.  If one uses the default setting of the package micrOMEGAs \cite{micrOMEGAs-1,micrOMEGAs-2,micrOMEGAs-3} for the nucleon form factors, i.e., $\sigma_{\pi N} = 34~{\rm MeV}$ and $\sigma_0 = 42~{\rm MeV}$~\cite{Ellis:2008hf}, in calculating $\sigma^{\rm SI}_{\tilde{\nu}_1-p}$, then $F_u^{p} \simeq 0.15$ and  $F_d^{p} \simeq 0.14$.
Alternatively, if one takes  $\sigma_{\pi N} = 59~{\rm MeV}$ and $\sigma_0 = 57~{\rm MeV}$ that were obtained in \cite{Alarcon:2011zs,Ren:2014vea,Ling:2017jyz} and \cite{Alarcon:2012nr}, respectively,
$F_u^{p} \simeq 0.16$ and  $F_d^{p} \simeq 0.13$.
This shows that different choices of the $\sigma_{\pi N}$ and $\sigma_0$ can induce uncertainties of ${\cal{O}} (10\%)$ in $F_u^{p}$ and $F_d^{p}$, and this does not drastically change the
cross-section. In addition, with the default setting, $F_u^{n} \simeq 0.15$ and  $F_d^{n} \simeq 0.14$. This implies the relation $\sigma^{\rm SI}_{\tilde{\nu}_1-p} \simeq \sigma^{\rm SI}_{\tilde{\nu}_1-n}$ for the
Higgs-mediated scattering.

The expressions of $a_{u_i}$ and $a_{d_i}$ reveal the following important features of the scattering:
\begin{itemize}
\item Although the SD cross-section vanishes, the SI cross-section depends not only on the parameters in the Higgs sector but also on those in the sneutrino sector, which include $\lambda_\nu$, $A_{\lambda_\nu}$, and $m_{\tilde{\nu}}$ for the Type-I NMSSM and $\lambda_\nu$, $Y_\nu$, $A_{\lambda_\nu}$, $A_\nu$, $m_{\tilde{\nu}}$, and $m_{\tilde{x}}$ for the ISS-NMSSM. This feature provides the theories (especially the ISS-NMSSM) with a great deal of freedom to be consistent with the experimental results. In particular, a small $|\mu|$ no longer enhance the cross-section, and it is quite often that, after fixing the parameters in the Higgs sector, an experimentally allowed DM candidate can be predicted by only adjusting the inputs of the sneutrino sector. In contrast, in the MSSM or the NMSSM with the lightest neutralino being a DM candidate, the SI and SD cross-sections rely only on the DM mass and the parameters in the Higgs sector. Due to different dependencies of the cross-sections on the parameters and also due to the constraints of the LHC experiments on the parameters, it is not easy to suppress the two cross-sections simultaneously in the light higgsino case~\cite{Huang:2014xua,Crivellin:2015bva,Carena:2018nlf,Badziak:2015exr,Badziak:2017uto}.
\item Each of the $h_i$ contributions to the SI cross-section is naturally suppressed. Explicitly speaking, the mass of the ${\rm Re}[H_d^0]$-dominated Higgs particle is usually at the TeV order, so its contribution is suppressed by the large mass. The ${\rm Re}[H_u^0]$-dominated scalar corresponds to the SM-like Higgs boson and its coupling with $\tilde{\nu}_1$ is suppressed by the factor $\lambda \lambda_\nu \cos\beta$ and the small $Y_\nu$. Moreover, as far as the ISS-NMSSM is concerned,  the accidental cancellation between the different terms in $C_{\tilde{\nu}_1^\ast \tilde{\nu}_1 {\rm Re}[H_u^0]}$ can further suppress the coupling. In most cases, the contribution from the singlet-dominated scalar is the most important, but it vanishes if there is no singlet-doublet mixing in the CP-even Higgs sector.
\end{itemize}
Due to these features, the extensions can easily satisfy the constraints of the DM DD experiments, and this was proven by the Bayesian analysis of the Type-I NMSSM~\cite{Cao:2018iyk}.

To further illustrate the behavior of the scattering rate, we consider a special case where $C_{\tilde{\nu}_1^\ast \tilde{\nu}_1 H_d} = C_{\tilde{\nu}_1^\ast \tilde{\nu}_1 H_u} = 0$ and $m_{H^\pm} \gtrsim 1 {\rm TeV}$.  We first integrate out the heavy doublet Higgs field so that the CP-even Higgs sector contains only the SM Higgs field ${\rm \sin \beta} {\rm Re}[H_u^0] + {\rm \cos \beta} {\rm Re}[H_d^0]$ and the singlet field ${\rm Re}[S]$. We then calculate the scattering amplitude by the mass insertion method.  The result takes the following form
\begin{eqnarray}
\sum_i a_{q_i} &=& -\frac{g}{8 m_W} \frac{C_{\tilde{\nu}_1^\ast \tilde{\nu}_1 S}}{m_{\tilde{\nu}_1}} \times \frac{(m_{h_2}^2 - m_{h_1}^2) \sin \theta \cos \theta}{m_{h_1}^2 m_{h_2}^2} \nonumber \\
   & \simeq & -\frac{g}{8 m_W} \frac{C_{\tilde{\nu}_1^\ast \tilde{\nu}_1 S}}{m_{\tilde{\nu}_1}} \times \frac{1}{ (125~{\rm GeV})^2} \times \delta \times \sin \theta \cos \theta,  \label{SI-expression}
   \end{eqnarray}
where $\delta$ denotes the splitting between $m_{h_1}^2$ and $m_{h_2}^2$ normalized by the squared mass of the singlet dominated Higgs boson, and $\theta$ is the mixing
angle of the SM Higgs field and the singlet field.
This formula indicates that, besides reducing $C_{\tilde{\nu}_1^\ast \tilde{\nu}_1 S}$ in the specific parameter space of the sneutrino sector, a small mixing angle obtained by adjusting the parameters in the Higgs sector can suppress the scattering. This small mixing, however, is favored by the Higgs data at the LHC.

At first glance, it appears that the formula in Eq.~(\ref{SI-expression}) may be applied to the NMSSM with the singlino-dominated neutralino being the DM candidate by the replacement $C_{\tilde{\nu}_1^\ast \tilde{\nu}_1 S}/m_{\tilde{\nu}_1} \to C_{\tilde{\chi}_1^0 \tilde{\chi}_1^0 S} \simeq \kappa$. This speculation is incorrect because, in order to obtain the correct density, the neutralino must contain sizable higgsino components, and this will induce the direct coupling of the neutralino with the SM Higgs field. As a result, the scattering cross-section is quite large, even for the case of $\sin \theta = 0$ (this is evident from the fact that $\lambda > 2 \kappa$, and it is demonstrated in the example of Case II in the Appendix). In the seesaw extensions of the NMSSM, however, the sneutrino DM may naturally correspond to an almost pure singlet field. In this case, its coupling with the SM Higgs field emerges only after electroweak symmetry breaking and is suppressed by the factors $\lambda \lambda_\nu v_d$ and $Y_\nu^2 v_u$.
This is a significant difference between these the extensions and the NMSSM.

Before concluding the introduction of the theories, we point out that DM physics in the $B_{\mu_X} = 0$ case of the ISS-NMSSM
is slightly different from the previous description in two aspects. One is that the sneutrino DM corresponds to a complex field,
and its anti-particle also acts as a DM candidate with an equal contribution to the relic density. As such,
this case is actually a two-component DM theory. The other is that the $Z$-boson can mediate the elastic 
scattering of the DM with nucleons, and consequently, it contributes to
the SI cross-section. Since  the total SI cross-section in such a theory is obtained by averaging over the  $ \tilde{\nu}_1 N$
and  $\tilde{\nu}_1^\ast N$ scatterings and the interferences between the $Z$ and the Higgs exchange diagrams for the two scatterings have opposite signs~\cite{Dumont:2012ee}, the SI cross-section can be written as~\cite{Arina:2007tm}
\begin{eqnarray}
\sigma_N^{\rm SI} \equiv \frac{\sigma_{\tilde{\nu}_1-N}^{\rm SI} + \sigma_{\tilde{\nu}_1^\ast - N}^{\rm SI} }{2}  =  \sigma_N^h + \sigma_N^Z,
\end{eqnarray}
where $\sigma_N^h$ is the same as before, and the $Z$-mediated contributions are given by
\begin{eqnarray}
\sigma_n^Z \equiv \frac{G_F^2 V_{11}^4}{2 \pi} \frac{m_n^2}{(1 + m_n/m_{\tilde{\nu}_1})^2}, \quad \sigma_p^Z \equiv \frac{G_F^2 V_{11}^4 (4 \sin^2 \theta_W - 1)^2}{2 \pi} \frac{m_p^2}{(1 + m_p/m_{\tilde{\nu}_1})^2},
\end{eqnarray}
where $G_F$ denotes the Fermi constant and $\theta_W$ is the weak angle. In this case, $\sigma_n^{\rm SI}$ may differ significantly from $\sigma_p^{\rm SI}$. Consequently, the effective cross-section of  the coherent scattering between the DMs and xenon nucleus (defined as the averaged cross-section $\sigma^{\rm SI}_{\tilde{\nu}_1-Xe}/A^2$, where $A$ denotes the mass number of the xenon nucleus) is given by the following general form:
\begin{eqnarray}
\sigma_{\rm eff}^{\rm SI} = 0.169 \sigma^{\rm SI}_p  +  0.347 \sigma^{\rm SI}_n  + 0.484 \sqrt{ \sigma^{\rm SI}_p  \sigma^{\rm SI}_n },
\end{eqnarray}
where the three coefficients on the right side of the equation are obtained by considering the abundance
of different xenon isotopes in nature. This effective cross-section has the property $\sigma_{\rm eff}^{\rm SI} = \sigma_N^{\rm SI}$
if $\sigma_p^{\rm SI}$ and $\sigma_n^{\rm SI}$ are equal, and it can be compared directly with the bound of the PandaX-II and
XENON-1T experiments if the SD scattering cross-section is negligible~\cite{Dumont:2012ee}.

Using the micrOMEGAs code~\cite{micrOMEGAs-1,micrOMEGAs-2,micrOMEGAs-3}, we confirmed that for $B_{\mu_X} \lesssim 200~{\rm GeV^2}$,
the DM observables, such as its relic density and its current annihilation rate $\langle \sigma v \rangle_0$, are insensitive to
the value of $B_{\mu_X}$~\cite{Cao:2017cjf}. However, as far as the SI cross-section is concerned, the predictions for the $B_{\mu_X} = 0$ and  $B_{\mu_X} \neq 0$ cases
differ significantly due to the reason discussed above. Numerically speaking, we found that the DM DD experiments require $|V_{11}| \lesssim 0.01$ for the $B_{\mu_X} = 0$ case,
since the $Z$-mediated contribution is usually much larger than the Higgs-mediated contribution for sizable $V_{11}$~\cite{Kakizaki:2016dza}. In contrast,  $|V_{11}|$ may be as large as 0.1 for the $B_{\mu_X} \neq 0$ case (we will present the results in our forthcoming work). We also confirmed that, when building the model file for the micrOMEGAs with the package SARAH~\cite{sarah-1,sarah-2,sarah-3}, taking $\tilde{\nu}_1$ as a complex field (corresponding to $B_{\mu_X} = 0$) or as a real field (corresponding to $B_{\mu_X} \neq 0$) significantly affects the time cost for calculating the relic density: in the $B_{\mu_X} \neq 0$ case, more Feynman diagrams must be calculated, and consequently, the computation is quite time expensive.

Throughout this work, we neglect the $Z$-mediated contribution by setting $B_{\mu_X} = 100~{\rm GeV^2}$.

\begin{table}[t]
\begin{center}
\begin{tabular}{|c|c|c|c|c|c|}
  \hline
   parameter& value & parameter & value& parameter& value \\ \hline
  $\tan \beta $ & 19.24 & $\lambda$ & 0.16 & $\kappa$ & 0.11  \\
  $A_\lambda$ & 1785.0 ${\rm GeV}$& $A_\kappa$ & -304.6 ${\rm GeV}$& $\mu $ & 147.7 ${\rm GeV}$ \\
  $m_{\tilde{q}}$ & 2000 ${\rm GeV}$ & $m_{\tilde{l}}$ & 2000 ${\rm GeV}$  & $A_{t}$ & 1354.7 ${\rm GeV}$\\
  $M_1$ & 2000 ${\rm GeV}$ & $M_2$ & 2000 ${\rm GeV}$ & $M_3$ & 5000 ${\rm GeV}$\\
  $m_{h_1}$ & 96.1 ${\rm GeV}$ & $m_{h_2}$ & 124.6 ${\rm GeV}$ & $m_{h_3}$ & 2332.9 ${\rm GeV}$ \\
  $m_{A_1}$ & 302.3 ${\rm GeV}$ & $m_{A_2}$ & 2332.8 ${\rm GeV}$ & $m_{{\tilde{\chi}}_1^0}$& 145.1 ${\rm GeV}$\\
  $m_{{\tilde{\chi}}_2^0}$ & 155.8 ${\rm GeV}$ & $m_{{\tilde{\chi}}_1^{\pm}}$ & 152.9 ${\rm GeV}$ & $m_{{\tilde{\chi}}_2^{\pm}}$ & 2024.6 ${\rm GeV}$\\
  $Z_{11}$ & -0.01 & $Z_{12}$ & -0.39 & $Z_{13}$ & 0.92\\
  $Z_{21}$ & 0.05 & $Z_{22}$ & 0.92& $Z_{23}$ & 0.39\\
  $Z_{31}$ & 0.99 & $Z_{32}$ & -0.05& $Z_{33}$ & -0.01\\
  \hline
\end{tabular}
\end {center}
\caption{A special configuration of the Higgs sector and the default setting of the other unimportant parameters,
which corresponds to the benchmark setting of Region I in~\cite{Cao:2019ofo}.
In this configuration, $h_2$ acts as the SM-like Higgs boson, $h_1$ and $A_1$ are singlet-dominated scalars, $\tilde{\chi}_{1,2}^0$ and
$\tilde{\chi}_1^\pm$ are higgsino-dominated electroweakinos, and $Z_{ij}$ for $i,j=1,2,3$ are the elements of the rotation that diagonalize
the mass matrix of the CP-even Higgs bosons in the basis $({\rm Re}[H_d^0], {\rm Re}[H_u^0], {\rm Re}[S])$. All of the masses and $Z_{ij}$ in this table
are obtained by the setting $Y_\nu = \lambda_\nu=0$. Non-zero $Y_\nu$ and $\lambda_\nu$ may slightly alter their values due to sneutrino loop effects.
In addition, the properties of $h_2$ are consistent with the Higgs results from the recent ALTAS analysis, which are based on $80~{\rm fb^{-1}}$ data at the 13~TeV LHC~\cite{ATLAS:2019slw}. }
\label{table2}
\end{table}

\section{Numerical results} \label{Section-results}

In this section, we compare how the sneutrino DM in the two extensions stays consistent with the bound of the XENON-1T experiment on the scattering cross-section under the premise of predicting the correct density and the photon spectrum of the DM annihilation in dwarf galaxies compatible with the Fermi-LAT observation. We choose the  benchmark-setting of Region I in~\cite{Cao:2019ofo} for the parameters in the Higgs sector with detailed information presented in Table \ref{table2}. This setting predicts $m_{h_1} \simeq 96~{\rm GeV}$, a TeV magnitude $v_s$ ($v_s \equiv \sqrt{2} \mu/\lambda = 1273.5~{\rm GeV}$), and large singlet-doublet Higgs mixing. Thus, the DM-nucleon scattering may be quite large by the expression of the $C_{\tilde{\nu}_1 \tilde{\nu}_1 S}$ and the formula in Eq.~(\ref{SI-expression}).
We emphasize that, although the setting is consistent with the latest data of the LHC related to the discovered Higgs boson and the search for extra bosons at the LEP and  LHC~\cite{Cao:2019ofo}, it is actually a rare case since the mixing angle is usually small in the broad parameter space of the NMSSM after considering the data. However, studying this extreme case is very helpful for improving our understanding of the scattering. In particular, it may reveal the mechanisms that keep the theories to be consistent with the tight XENON-1T bound.

The procedure in our study is as follows. We first construct the likelihood function for the DM physics and perform sophisticated scans over the parameters of the sneutrino sector for either theory by requiring the lightest sneutrino as a DM candidate. We then plot the map of the profile likelihood (PL)~\cite{Fowlie:2016hew,Cao:2018iyk}
\footnote{In frequentist statistics, the PL for a likelihood function $\mathcal{L}$ represents the most significant likelihood value in a specific parameter space. With the two dimensional (2D) PL as an example, it is defined by
	\begin{eqnarray}
 \mathcal{L}(\Theta_A,\Theta_B)=\mathop{\max}_{\Theta_1,\cdots,\Theta_{A-1},\Theta_{A+1},\cdots, \Theta_{B-1}, \Theta_{B+1},\cdots}\mathcal{L} (\Theta). \nonumber
	\end{eqnarray}
where $\Theta = (\Theta_A,\Theta_B,\cdots)$ is a set of parameters that $\mathcal{L}$ depends on, and one obtains the maximization by changing parameters other than $\Theta_A$ and $\Theta_B$. }
in different two-dimensional planes to illustrate their features and determine the underlying physics. The likelihood function we adopt is expressed as follows:
\begin{eqnarray}
	\mathcal{L}_{\rm DM} = \mathcal{L}_{\Omega_{\tilde{\nu}_1}} \times  \mathcal{L}_{\rm DD} \times \mathcal{L}_{\rm ID},   \label{DM-profile}
\end{eqnarray}
where $\mathcal{L}_{\Omega_{\tilde{\nu}_1}}$, $\mathcal{L}_{\rm DD}$, and $\mathcal{L}_{\rm ID}$ account for the relic density, the XENON-1T experiment and the Fermi-LAT observation of dwarf galaxies, respectively. Their expressions are as follows:
\begin{itemize}
\item $\mathcal{L}_{\Omega_{\tilde{\nu}_1}}$ is Gaussian distributed, i.e.,
\begin{equation}
\mathcal{L}_{\Omega_{\tilde{\nu}_1}}=e^{-\frac{[\Omega_{\rm th}-\Omega_{\rm obs}]^2}{2\sigma^2}},
\end{equation}
where $\Omega_{th}$ denotes the theoretical prediction of the density $\Omega_{\tilde{\nu}_1} h^2$, $\Omega_{\rm obs}=0.120$ represents
its experimental central value~\cite{Aghanim:2018eyx}, and $\sigma = 0.1 \times \Omega_{\rm obs}$ is the total (including both theoretical and experimental) uncertainty of the density.
\item $\mathcal{L}_{\rm DD}$ takes a Gaussian form with a mean value of zero~\cite{Matsumoto:2016hbs}:
\begin{equation}
\mathcal{L}_{\rm DD}=e^{-\frac{\sigma^2_{\tilde{\nu}_1-p}}{2\delta_{\sigma}^2}},
\end{equation}
where $\sigma_{\tilde{\nu}_1-p}$ denotes the theoretical prediction of the DM-proton scattering rate, $\delta_{\sigma}$ is evaluated by $\delta_{\sigma}^2 = UL_\sigma^2/1.64^2 + (0.2\sigma_{\tilde{\nu}_1-p})^2$, $UL_\sigma$ denotes the upper limit of the latest XENON-1T results on the scattering cross-section at a $90\%$ C.L.~\cite{Aprile:2018dbl}, and $0.2 \sigma_{\tilde{\nu}_1-p}$ parameterizes the theoretical uncertainty of $\sigma_{\tilde{\nu}_1-p}$.
\item $\mathcal{L}_{\rm ID}$ is calculated by the likelihood function proposed in~\cite{Likelihood-dSph,Zhou} with the data of the Fermi-LAT collaboration presented in~\cite{Ackermann:2015zua,Fermi-Web}.
\end{itemize}

In our study, we utilized the package \textsf{SARAH-4.11.0}~\cite{sarah-1,sarah-2,sarah-3} to build the models, the \textsf{SPheno-4.0.3} code \cite{spheno}
to generate the particle spectrum, and the package \textsf{MicrOMEGAs 4.3.4}~\cite{Belanger:2013oya,micrOMEGAs-1,micrOMEGAs-3}
to calculate the DM observables. We set the soft breaking masses for the first two generation sneutrino fields at $2~{\rm TeV}$ throughout this work.

\begin{figure*}[t]
		\centering
		\resizebox{0.9\textwidth}{!}{
        \includegraphics[width=0.90\textwidth]{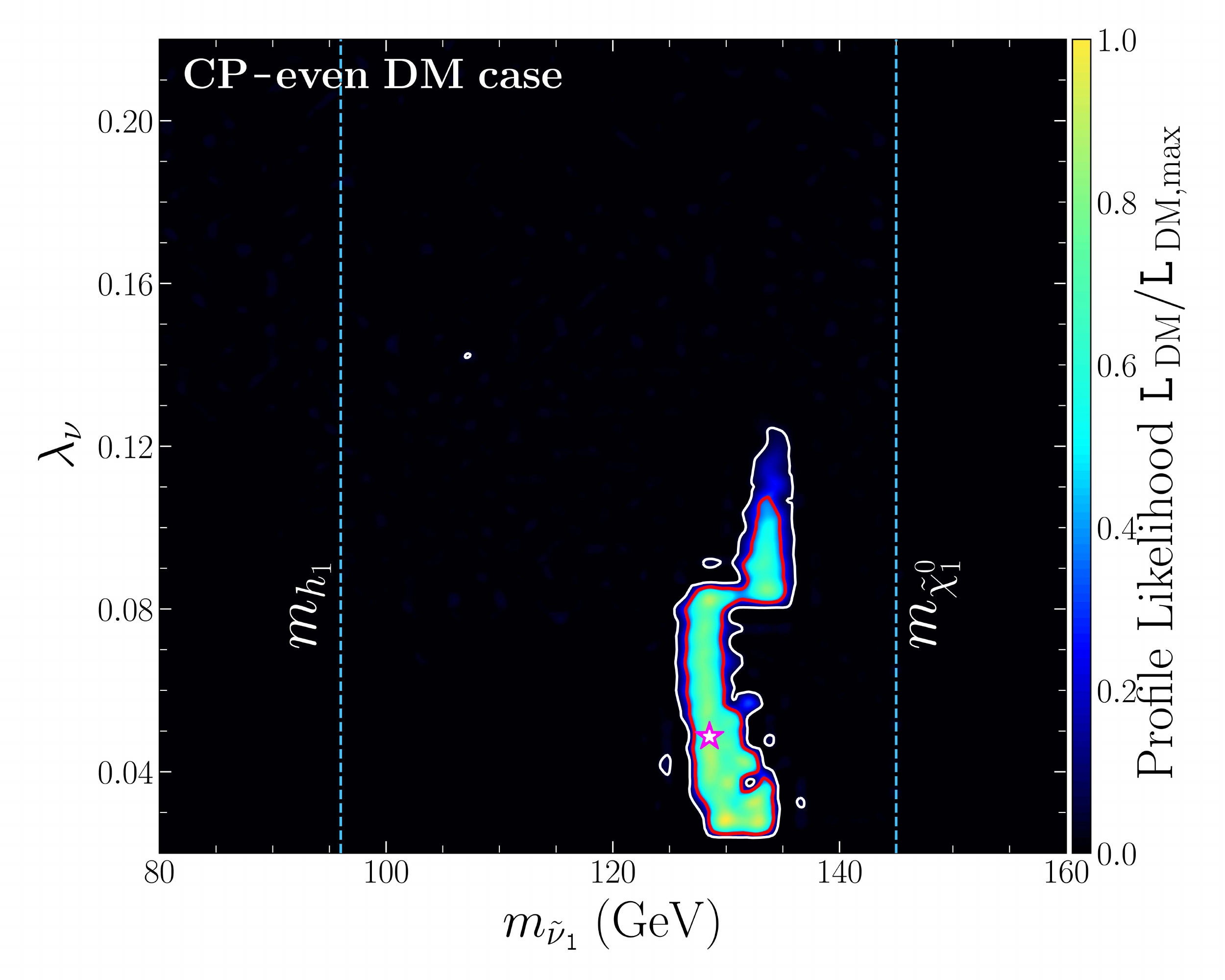}
        \includegraphics[width=0.90\textwidth]{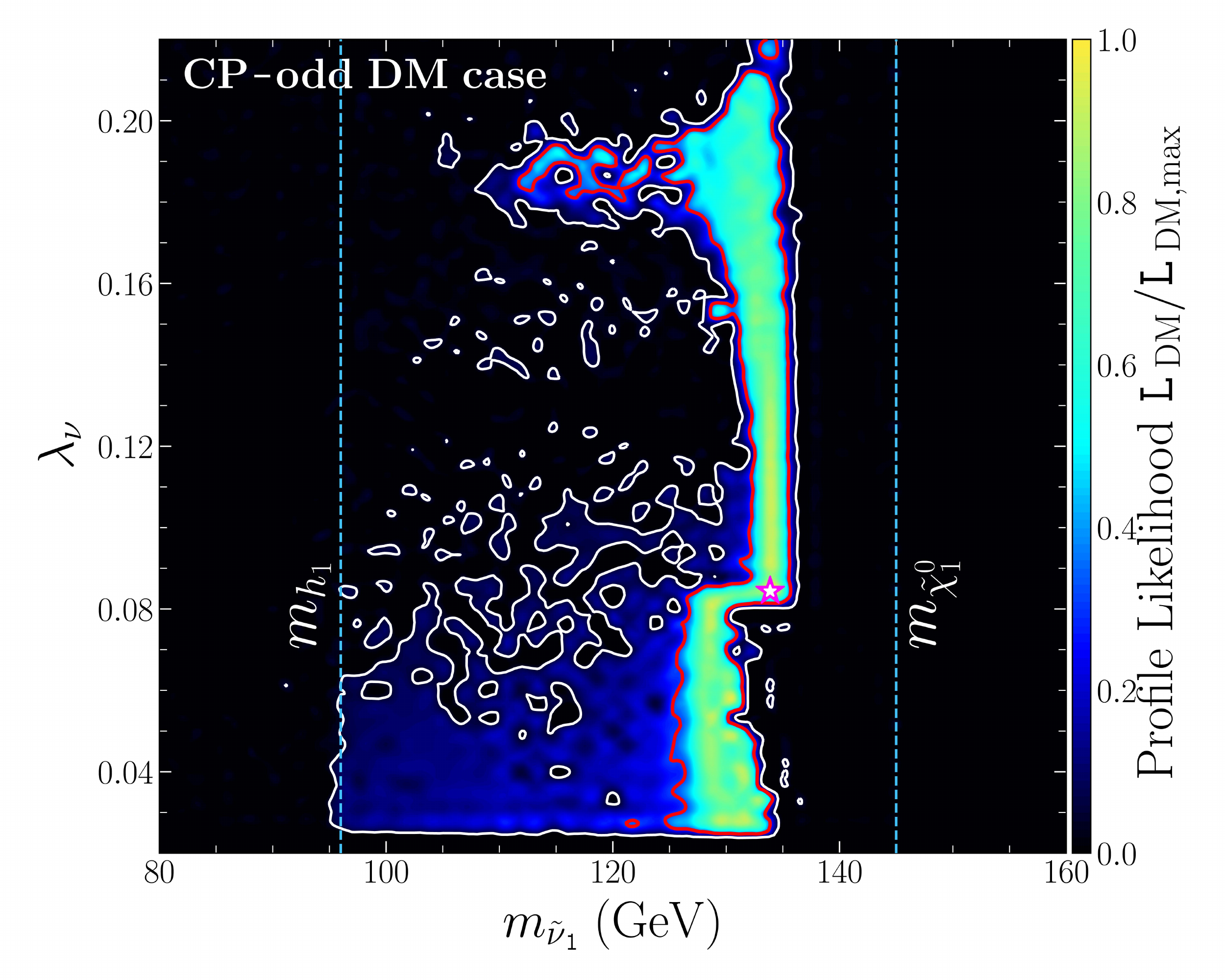}
        }

\vspace{0.1cm}

		\resizebox{0.9\textwidth}{!}{
        \includegraphics[width=0.90\textwidth]{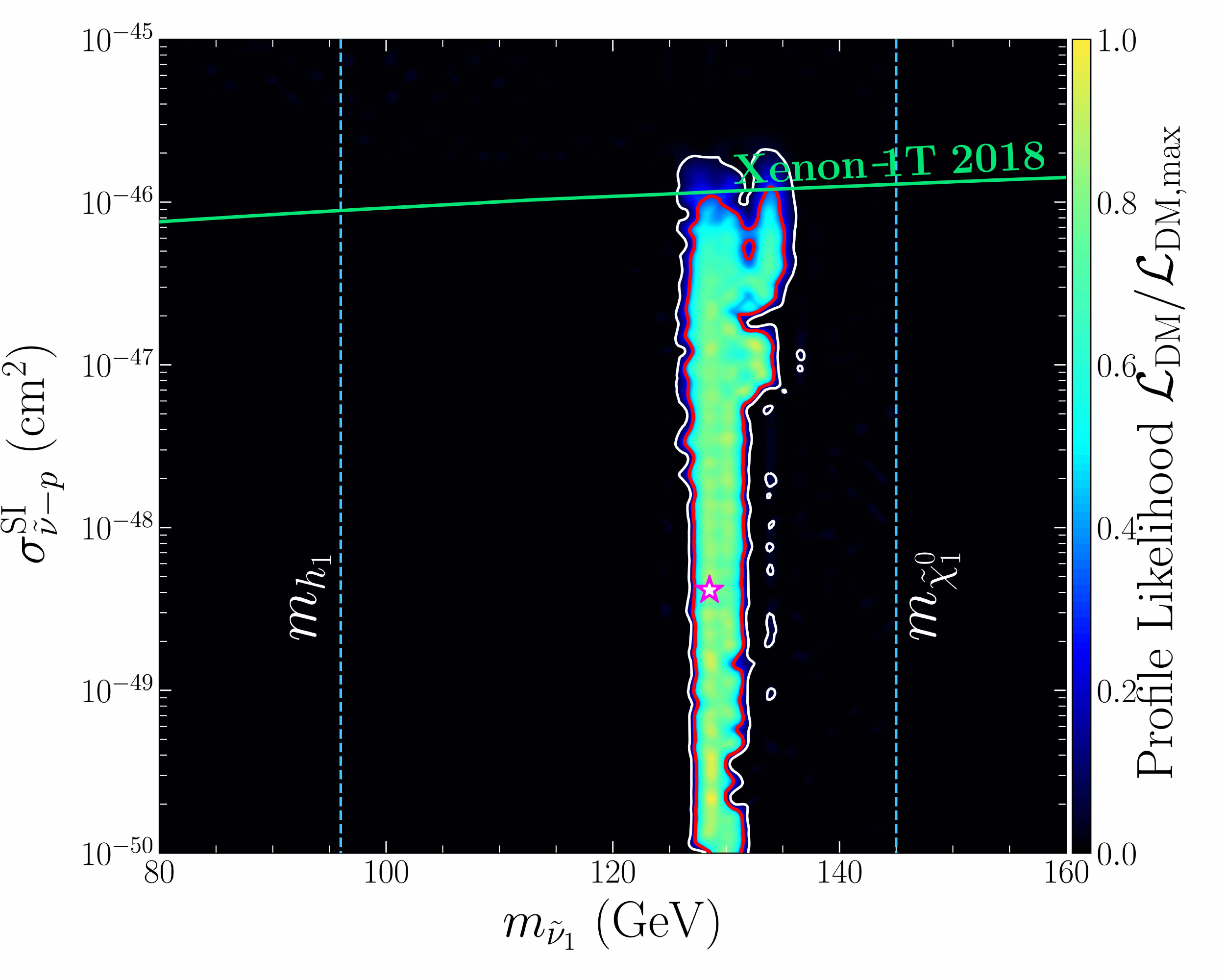}
        \includegraphics[width=0.90\textwidth]{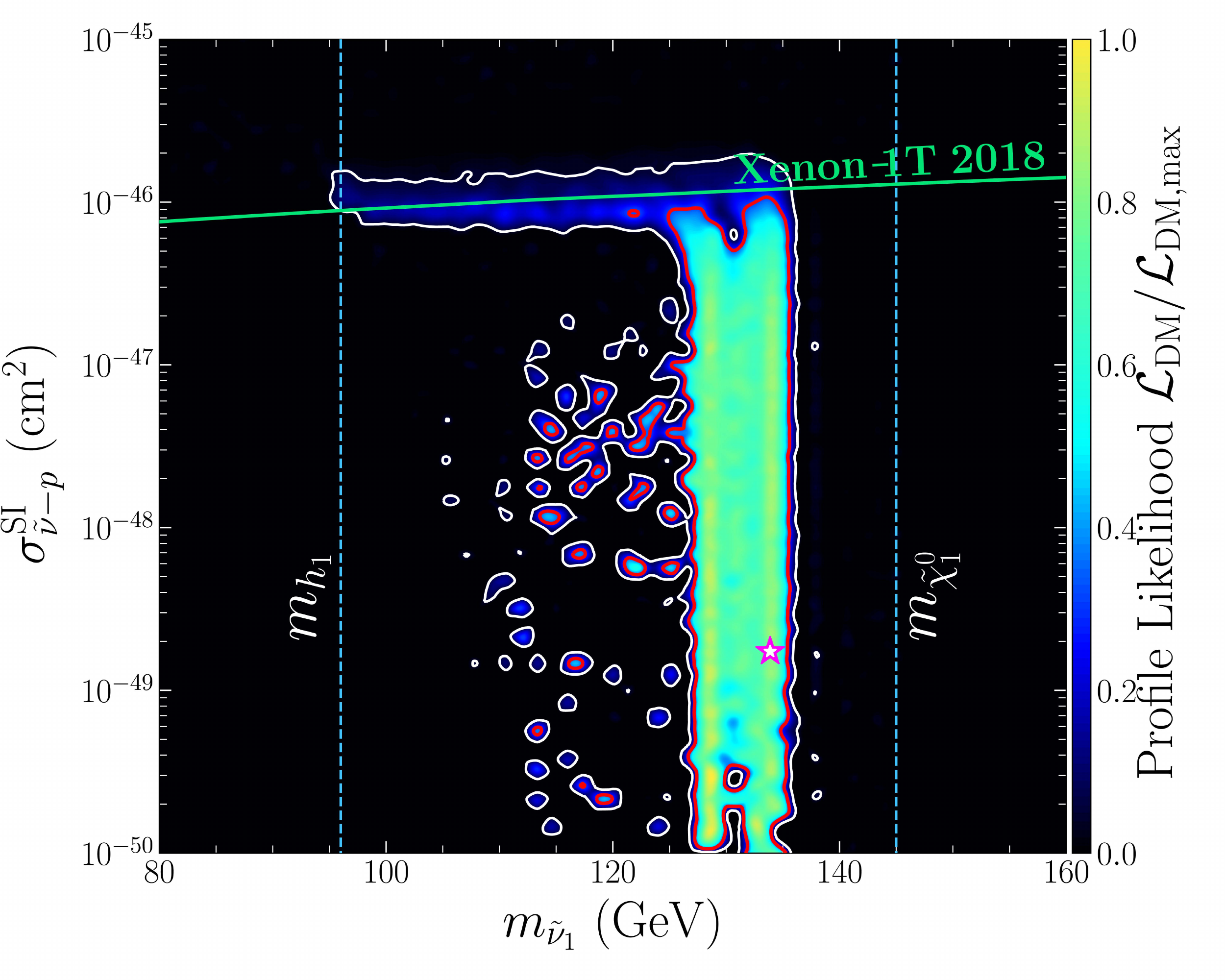}
        }
       \vspace{-0.3cm}
        \caption{Two-dimensional profile likelihoods of the function $\mathcal{L}$ in Eq.~(\ref{DM-profile}), which are projected onto the $\lambda_\nu-m_{\tilde{\nu}_1}$ and $\sigma_{\tilde{\nu}_1-p}^{\rm SI}-m_{\tilde{\nu}_1}$ planes in the framework of the Type-I NMSSM, respectively. The left panels are the results for the CP-even DM case, and the right panels are for the CP-odd DM case. Since $\chi^2_{\rm min} \simeq 0$ for the best point (marked with star symbol in the figure),
        the boundary for $1 \sigma$ confidence interval (red solid line) and that for $2\sigma$ confidence interval (white solid line) correspond to $\chi^2 \simeq 2.3$ and $\chi^2 \simeq 6.18$, respectively. This figure reflects the preference of the DM measurements on the parameter space of the Type-I NMSSM. \label{fig1} }
\end{figure*}	

\subsection{Features of sneutrino DM in Type-I NMSSM}

We performed two independent scans over the following parameter space of the Type-I NMSSM\footnote{We note that the SM-like Higgs boson may decay into a massive neutrino pair if it is lighter than about $60~{\rm GeV}$, and the branching ratio may be significantly large since the boson in the benchmark setting contains a sizable singlet Higgs component. To avoid such a possibility, we require $\lambda_\nu \gtrsim 0.025$, so the decay is kinematically forbidden. }
\begin{eqnarray}
0 < m_{\tilde{\nu}} < 200~{\rm GeV},\quad 0.025 < \lambda_\nu < 0.5, \quad |A_{\lambda_\nu} | < 1~{\rm TeV},   \label{DM-parameter}
\end{eqnarray}
with the MultiNest algorithm~\cite{Feroz:2008xx,Feroz:2013hea} by taking the prior probability density function (PDF) as uniformly distributed and $\tilde{\nu}_1$ as a DM candidate to be CP-even and CP-odd, respectively. With the samples obtained in the scan, we show the map of the PL for the function $\mathcal{L}_{\rm DM}$ on different planes in Fig.~\ref{fig1}. From the results of the left panels for the CP-even case, following facts are obtained
\begin{itemize}
\item $m_{\tilde{\nu}_1}$ is concentrated on the range from $125$ to $135~{\rm GeV}$, which is close to $m_{\tilde{\chi}_1^0}$. Since $\lambda_\nu$ is small, the annihilations $\tilde{\nu}_1 \tilde{\nu}_1 \to h_1 h_1, h_1 h_2, h_2 h_2$ are unimportant (discussed below), and the sneutrino DM mainly co-annihilated with the higgsino-dominated electroweakinos to achieve its measured density. In this case, the relic density is much more sensitive to the splitting between $m_{\tilde{\nu}_1}$ and $\mu$ than to $\lambda_\nu$.
\item $\lambda_\nu$ in the $2 \sigma$ confidence interval (CI) is upper bounded, e.g., $\lambda_\nu \lesssim 0.09$ for $m_{\tilde{\nu}_1} \simeq 128~{\rm GeV}$ and $\lambda_\nu \lesssim 0.12$ for $m_{\tilde{\nu}_1} \simeq 134~{\rm GeV}$. This feature is mainly due to the constraints from the XENON-1T experiment, which can be understood as follows.
     First, for the parameters in Table \ref{table2} (in particular the remarkable feature that both the ratio $\mu/\lambda$ and the singlet-doublet Higgs mixing are quite large), the $C_{\tilde{\nu}_1\tilde{\nu}_1 h_1}$ and $C_{\tilde{\nu}_1\tilde{\nu}_1 h_2}$ couplings in Eq.~(\ref{Csnn}) are contributed to mainly by the terms in the second brackets. Thus, one can conclude from Eq.~(\ref{SI-expression}) that
    \begin{eqnarray}
    \sum_i a_{q_i} \simeq  -\frac{g}{8 m_W m_{\tilde{\nu}_1}} \times \left \{ \frac{\sqrt{2}}{\lambda} \left(2\lambda_{\nu}^2 + \kappa\lambda_{\nu}\right) \mu  - \frac{\lambda_{\nu} A_{\lambda_\nu}}{\sqrt{2}} \right \} \times \frac{0.25}{ (125~{\rm GeV})^2}.  \label{SI-special case}
   \end{eqnarray}
   This approximation reflects that the SI cross-section is roughly proportional to $1/m_{\tilde{\nu}_1}^{2}$, and it increases monotonically with $\lambda_\nu$.  Second, the
   soft-breaking term of the Type-I NMSSM in Eq.~(\ref{superpotential}) and the sneutrino mass matrix in Eq.~(\ref{sneutrino_matrix}) show that the parameter $m_{\tilde{\nu}}$ does not introduce any interactions, and it is related to $m_{\tilde{\nu}_1}$ by
  \begin{eqnarray}
   m_{\tilde{\nu}_1}^2 & \simeq & m^2_{R\bar{R}} + 2 m^2_{RR} \nonumber \\
    &=& m^2_{\tilde{\nu}} + 2|\lambda_{\nu}v_s|^2+\frac{1}{2}|Y_{\nu}v_u|^2 + \lambda_{\nu}[\sqrt{2}A_{\lambda_{\nu}}v_s+(\kappa v^2_s-\lambda v_dv_u)^*].  \label{mass-relation}
   \end{eqnarray}
    As a result, one may substitute $m_{\tilde{\nu}}$ with $m_{\tilde{\nu}_1}$ as an input parameter. In this way, $m_{\tilde{\nu}_1}$ no longer relies on $\lambda_\nu$ and $A_{\lambda_{\nu}}$ except that, to make the CP-even sneutrino state as a DM candidate lighter than its CP-odd partner, $A_{\lambda_\nu}$ should be negative and satisfy $|A_{\lambda_\nu}| \gtrsim \kappa/\lambda \mu $ with $\kappa/\lambda \mu \simeq 100~{\rm GeV}$ in our parameters. However, a negative $A_{\lambda_\nu}$ ensures that the second term in the curl brackets of Eq.~(\ref{SI-special case}) always interferes constructively with the first term to strengthen the constraint of the DD experiment. Finally, we emphasize that a relaxed experimental bound on the scattering cross-section with the increase in $ m_{\tilde{\nu}_1}$ is also an important factor for determining the upper bound.

\item The annihilation $\tilde{\nu}_1 \tilde{\nu}_1 \to h_1 h_1$ for $m_{h_1} \gtrsim 96~{\rm GeV}$ is unable to account for the measured DM density. This is because the density requires $\lambda_\nu \simeq 0.17$ for the parameters in Table \ref{table2} if the channel is fully responsible for the density through the quartic scalar interactions (this conclusion was obtained by the formula of the relic density in~\cite{Chang:2013oia,Berlin:2014tja}). Such a large $\lambda_\nu$, however, is strongly disfavored by the XENON-1T experiment.
\end{itemize}

As for the upper left panel of Fig.~\ref{fig1}, it may appear that the dependence of the SI cross-section on the parameter $\lambda_\nu$ becomes a step function at $\lambda_\nu \simeq 0.09$. A proper understanding of the panel involves the concept of the two dimensional (2D) PL $\mathcal{L} (\lambda_\nu, m_{\tilde{\nu}_1})$,
which is defined by (footnote 5)
\begin{eqnarray}
\mathcal{L} (\lambda_\nu, m_{\tilde{\nu}_1}) = \max_{|A_{\lambda_\nu}| \leq 1~{\rm TeV}} \mathcal{L}_{DD} (\lambda_\nu, A_{\lambda_\nu}, m_{\tilde{\nu}_1}). \label{DM-profile1}
\end{eqnarray}
In plotting the panel, the maximization over $A_{\lambda_\nu}$ was obtained through the following procedure: we split the $\lambda_\nu-m_{\tilde{\nu}_1}$ plane into
$80 \times 80$ equal boxes (i.e., we divided each dimension of the plane by 80 equal bins), fitted the samples obtained in the scan into each box so that
the samples in each box corresponded roughly equal $\lambda_\nu$ and $m_{\tilde{\nu}_1}$ (but $A_{\lambda_\nu}$ might differ greatly), and finally  selected the maximum likelihood
value from the samples in each box as the PL value. It is obvious that $\mathcal{L} (\lambda_\nu, m_{\tilde{\nu}_1})$ reflects the preference of the theory on the parameters $\lambda_\nu$ and $m_{\tilde{\nu}_1}$, and for a given point in the $\lambda_\nu - m_{\tilde{\nu}_1}$ plane, its value represents the capability of the point to account for experimental data. Sequentially one can define the 2D CI as the region on the plane satisfying
\begin{eqnarray}
\chi^2 - \chi^2_{\rm min} \leq 6.18,
\end{eqnarray}
where $\chi^2 \equiv -2 \ln \mathcal{L} (\lambda_\nu, m_{\tilde{\nu}_1})$, and $\chi^2_{\rm min}$ is the $\chi^2$ value for the best sample (best point) obtained in the scan. In our study, $\chi^2_{\rm min} \simeq 0$ because the DM experimental data are independent and consistent with each other, and the Type-I NMSSM can explain the data well. With this knowledge, one can infer that the CIs are not necessarily contiguous~\cite{Fowlie:2016hew,Cao:2018iyk}. In fact, the panel actually reveals the following conclusions: the 2D CI is mainly located in two isolated parameter islands in the $\lambda_\nu - m_{\tilde{\nu}_1}$ plane, which are featured by $\lambda_\nu \lesssim 0.09$, $125~{\rm GeV} \lesssim  m_{\tilde{\nu}_1} \lesssim 135~{\rm GeV} $, and $\lambda_\nu \gtrsim 0.08$, $m_{\tilde{\nu}_1} \sim 134~{\rm GeV} $, respectively, and the two islands are connected at $m_{\tilde{\nu}_1} \sim 132~{\rm GeV}$ with $0.08 \leq \lambda_\nu \leq 0.09$. We confirmed that at the bridge, $A_{\lambda_\nu}$ may vary from $-260$ to $-410~{\rm GeV}$ to predict $\chi^2 \leq 6.18$.

Next, we concentrate on the CP-odd sneutrino DM case, where $a_{q_i}$ is given by (discussed below Eq.~(\ref{Csnn}))
    \begin{eqnarray}
    \sum_i a_{q_i} \simeq  -\frac{g}{8 m_W m_{\tilde{\nu}_1}} \times \left \{ \frac{\sqrt{2}}{\lambda} \left(2\lambda_{\nu}^2 - \kappa\lambda_{\nu}\right) \mu  + \frac{\lambda_{\nu} A_{\lambda_\nu}}{\sqrt{2}} \right \} \times \frac{0.25}{ (125~{\rm GeV})^2}.  \label{SI-special case1}
   \end{eqnarray}
Compared with the CP-even case, the results on the right panels of Fig.~\ref{fig1} show similar features except for three aspects. The first is that $\lambda_\nu$ in the co-annihilation region can take a much larger value than the CP-even case. The reason is that $A_{\lambda_\nu}$ in the CP-odd case may be either positive or negative, and consequently, the second term in the curly brackets of Eq.~(\ref{SI-special case1}) can interfere destructively with the first term to weaken the constraint of the DM DD experiment.
The second aspect is that the $1 \sigma$ CI in the $\lambda_\nu-m_{\tilde{\nu}_1}$ plane includes some separated islands with $m_{\tilde{\nu}_1} \lesssim 125~{\rm GeV}$ and $\lambda_\nu \simeq 0.17 $. For samples located in these islands, the annihilations $\tilde{\nu}_1 \tilde{\nu}_1 \to h_1 h_1, h_1 h_2$ can account for the measured density and the scattering cross-section can be consistent with the bound of the XENON-1T experiment due to the cancellation. The last aspect is that there is a broad $2\sigma$ CI characterized by $m_{\tilde{\nu}_1} \lesssim 125~{\rm GeV}$ and $\lambda_\nu \lesssim 0.11$. In this region, the correct DM density cannot be achieved by the co-annihilation and the annihilation into Higgs bosons, although the constraint from the DM DD experiment can be satisfied. This region is a compromise of the two constraints that maximizes the PL in Eq.~(\ref{DM-profile1}).

The information of the best point for the two cases is as follows:
\begin{itemize}
\item CP-even case: \\
$\lambda_\nu = 0.05$, $A_{\lambda_\nu} = -438~{\rm GeV}$, $m_{\tilde{\nu}} = 192.3~{\rm GeV}$, $\Omega_{th} = 0.120$, $\sigma_{\tilde{\nu}_1-p} = 4.1 \times 10^{-49}~{\rm cm^2}$,
$\Delta_{\Omega} = 42.2$, $\Delta_\sigma = 1.4$, $\bar{\chi}^2 \simeq 0$;
\item CP-odd case: \\
$\lambda_\nu = 0.08$, $A_{\lambda_\nu} = 58.6~{\rm GeV}$, $m_{\tilde{\nu}} = 138.2~{\rm GeV}$, $\Omega_{th} = 0.119$, $\sigma_{\tilde{\nu}_1-p} = 1.7 \times 10^{-49}~{\rm cm^2}$,
$\Delta_{\Omega} = 20.2$, $\Delta_\sigma = 1.8$, $\bar{\chi}^2 \simeq 0$;
\end{itemize}
where $\bar{\chi}^2 \equiv -2 \ln \mathcal{L}_{\rm DM} $, and the fine-tuning quantities $\Delta_{\Omega}$ and $\Delta_\sigma$ are defined by\footnote{In the definition of $\Delta_\sigma$, the unit of $\sigma_{\tilde{\nu}_1-p}$, i.e. $10^{-47} {\rm cm^2}$, represents its current experimental sensitivity. Since the relic density has been precisely measured while the scattering cross-section is only upper bounded, we chose different definitions of $\Delta_{\Omega}$ and $\Delta_\sigma$ to parameterize the theory's ability to account for the experimental results. }
\begin{eqnarray}
\Delta_{\Omega} \equiv {\rm Max}_i \left | \frac{\partial \left [ \Omega_{\rm th}/\Omega_{\rm obs} \right ]}{\partial \ln p_i} \right |, \quad \Delta_{\sigma} \equiv {\rm Max}_i \left | \frac{\partial \left [ \sigma_{\tilde{\nu}_1-p}/(10^{-47} {\rm cm^2}) \right ]}{\partial \ln p_i} \right |,
\end{eqnarray}
where $p_i$ denotes any of the input parameters of the theory. These quantities reflect the adaptation of the parameter point to be consistent with the experimental measurements. The larger they are, the more finely tuned the theory will be to coincide with the measurements. From our numerical results, we found that $\Omega_{th}$ is most sensitive to the mass splitting between $\tilde{\nu}_1$ and $\tilde{\chi}_1^0$, i.e., to $m_{\tilde{\nu}}$, $\lambda_\nu$, and $A_{\lambda_\nu}$. In contrast, $\sigma_{\tilde{\nu}_1-p}$ is sensitive only to $\lambda_\nu$ and $A_{\lambda_\nu}$. Given the values of $\Delta_{\Omega}$ and $\Delta_\sigma$ for the best points, we concluded that the Type-I NMSSM is natural in DM physics for either case. Moreover, we also calculated the Bayesian evidence $Z$ and found that $\ln Z = -8.98$ for the CP-even case and $\ln Z = -8.27$ for the CP-odd case. Since the Jeffrey's scale was only 0.71, we concluded that the theory does not show a strong preference of the CP-odd case over the CP-even case~\cite{Jeffreys,Feroz:2008wr}.

\begin{figure*}[t]
		\centering
        \resizebox{0.9\textwidth}{!}{
        \includegraphics[width=0.9\textwidth]{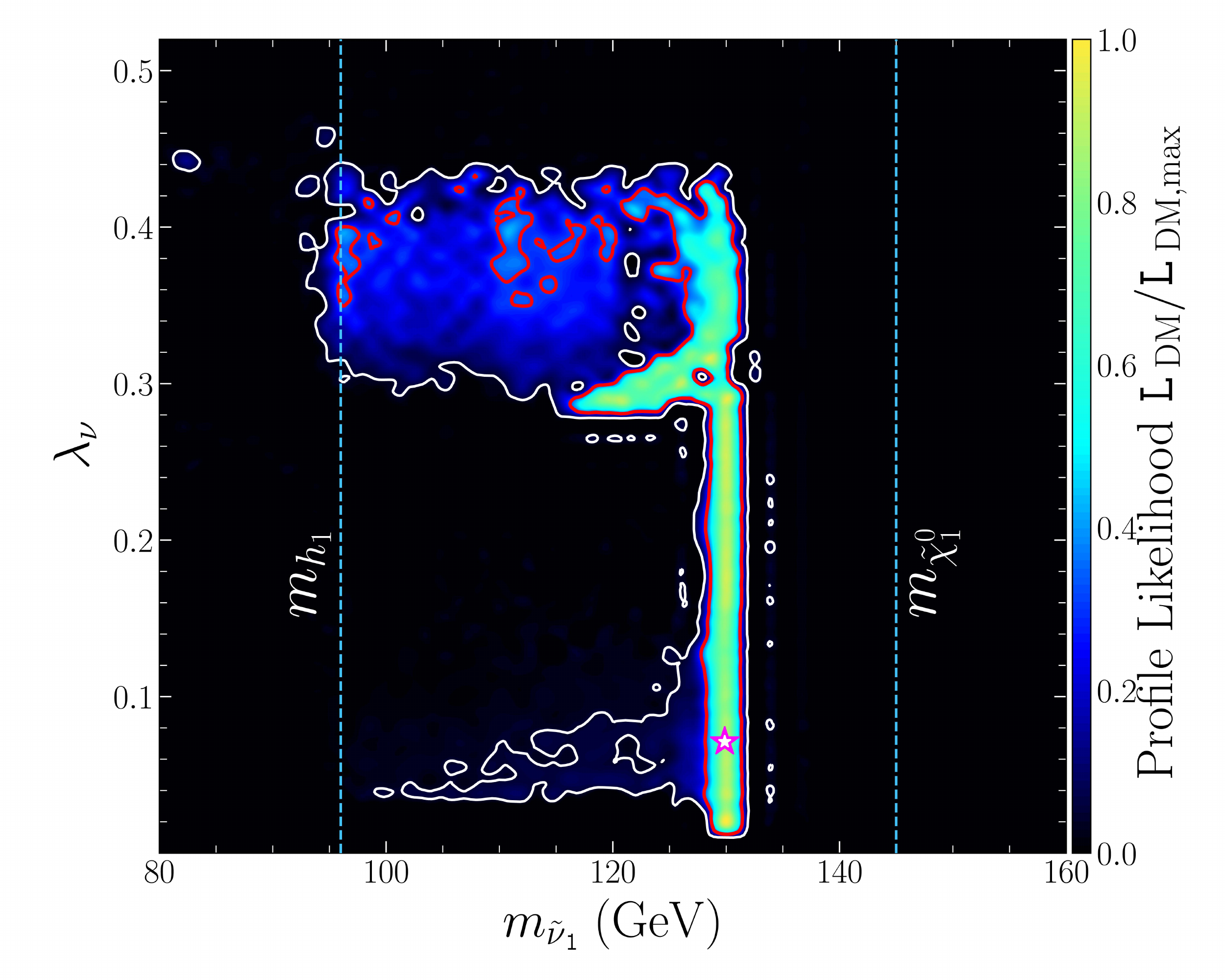}
        \includegraphics[width=0.9\textwidth]{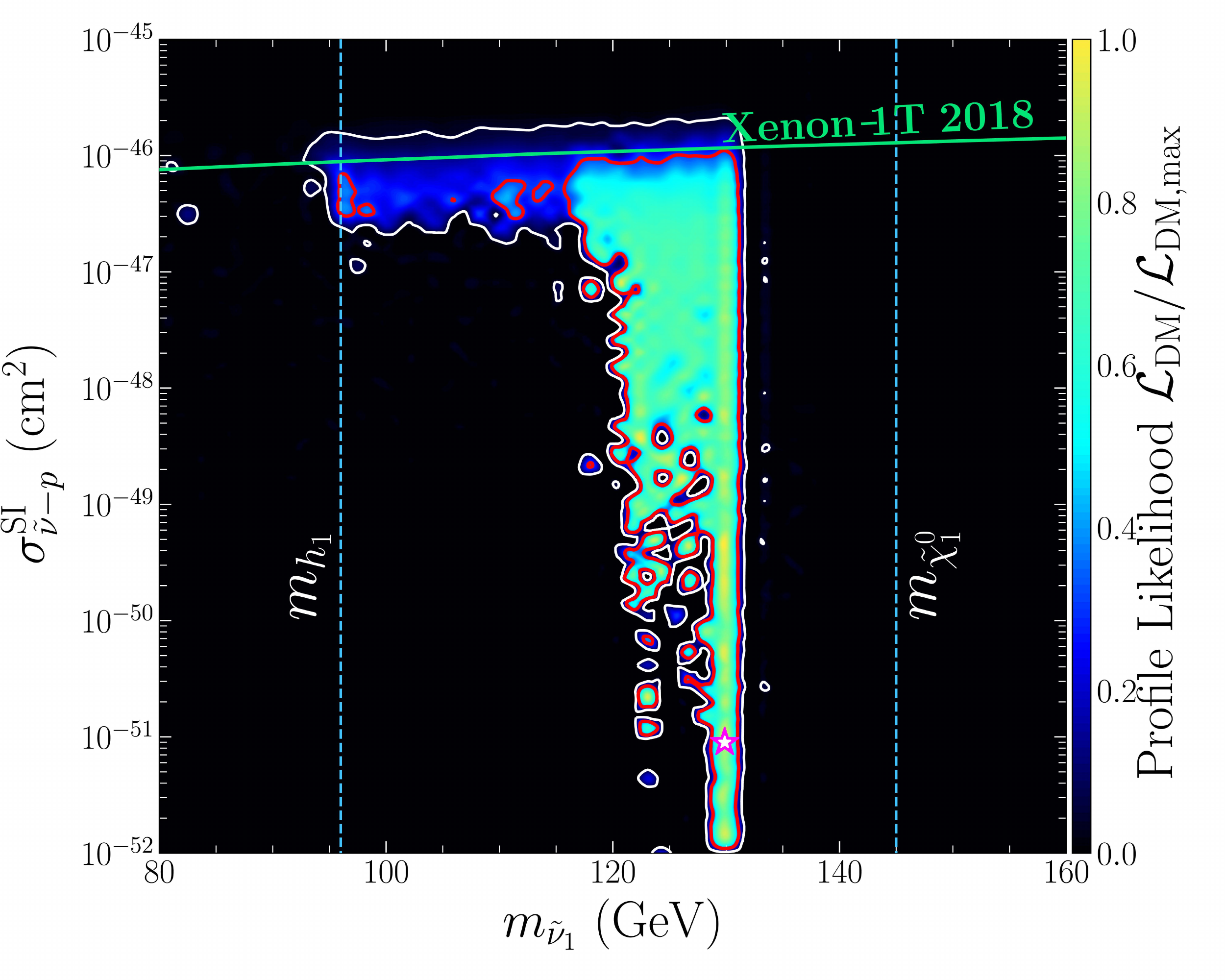}
        }
       \vspace{-0.6cm}
        \caption{The map for the profile likelihood of the function $\mathcal{L}_{\rm DM} $ in Eq.~(\ref{DM-profile}), which is plotted in the $\lambda_\nu-m_{\tilde{\nu}_1}$ and $\sigma_{\tilde{\nu}_1-p}^{\rm SI}-m_{\tilde{\nu}_1}$ planes in the framework of the ISS-NMSSM, respectively. Since $\chi^2_{\rm min} \simeq 0$ for the best point (marked with star symbol in the figure), the $1 \sigma$ boundary (red solid line) and the $2\sigma$ boundary (white line) correspond to $\chi^2 \simeq 2.3$ and $\chi^2 \simeq 6.18$, respectively. This figure reflects the preference of the DM measurements in the parameter space of the ISS-NMSSM. \label{fig2} }
\end{figure*}	

Before concluding this section, we emphasize that the $1\sigma$ CIs of our results for either case also include the regions characterized by $m_{\tilde{\nu}_1} \simeq m_{h_1}/2$ or by $m_{\tilde{\nu}_1} \simeq m_{h_2}/2$. In these regions, the DM was annihilated by a resonant Higgs boson to obtain its measured density. Quite similar to the co-annihilation case, $\lambda_\nu$ may be very small, and consequently, $\sigma_{\tilde{\nu}_1-p}$ can be as small as $10^{-50}~{\rm cm^2}$. These regions, however, have been excluded by the LHC search for the $2 \tau + E_{T}^{\rm Miss}$ signal, which is induced by the process $p p \to \tilde{\chi}_1^\pm \tilde{\chi}_1^\mp$, due to the large mass splitting between $\tilde{\chi}_1^\pm$ and $\tilde{\nu}_1$~\cite{Cao:2018iyk}. As such, we do not show them on the panels.

\begin{table*}[t]
	\begin{center}
\begin{tabular}{|c|c|c|c|c|c|}
\hline
 &  Point I &  Point II &  & Point I & Point II \\ \hline
$Y_\nu$ & 0.03 & 0.16  & $V_{11}$& 0 &  0 \\ \hline
$\lambda_\nu$ & 0.07  & 0.40   & $V_{12}$& -0.50& 0.72  \\ \hline
$A_\nu$ & -398.2~{\rm GeV}  & 393.0~{\rm GeV}  & $V_{13}$ & -0.87 &  -0.70 \\ \hline
$A_{\lambda_\nu}$ &  -372.1~{\rm GeV} & 344.1~{\rm GeV}  & $\Omega_{th}$  & 0.120& 0.125   \\ \hline
$m_{\tilde{\nu}}$ &  207.6~{\rm GeV} & 187.3~{\rm GeV}  & $\Delta_{\Omega}$  & 25 &  167.3 \\ \hline
$m_{\tilde{x}}$ &  153.3~{\rm GeV} & 213.1~{\rm GeV}  &$\sigma_{\tilde{\nu}_1-p}^{\rm SI}$ & $9.0 \times 10^{-52}~{\rm cm^2}$ &  $4.2 \times 10^{-47}~{\rm cm^2}$  \\ \hline
$m_{\tilde{\nu}_1}$ &  130.4~{\rm GeV} & 96.5~{\rm GeV}  & $\Delta_\sigma$ & 0.08 & 99.0  \\ \hline
\end{tabular}
\caption{Information of the benchmark points in the ISS-NMSSM.} \label{table3}
\end{center}
\end{table*}

\subsection{Features of sneutrino DM in ISS-NMSSM}

Similar to our approach for the Type-I NMSSM, we scanned the following parameter space of the ISS-NMSSM:
\begin{eqnarray}
0< Y_\nu, \lambda_\nu < 0.7, \quad 0 < m_{\tilde{\nu}}, m_{\tilde{x}} < 250~{\rm GeV}, \quad |A_\nu|, |A_{\lambda_\nu} | < 1{\rm TeV},
\end{eqnarray}
by taking the lightest sneutrino as a DM candidate and requiring that the Yukawa couplings $Y_\nu$ and $\lambda_\nu$ satisfy the unitary constraint in Eq.~(\ref{unitaryconstriants}). The results of the CIs are projected onto the $\lambda_\nu-m_{\tilde{\nu}_1}$ and $\lambda_\nu-\sigma_{\tilde{\nu}_1-p}$ planes in Fig.~\ref{fig2}. From this figure, one can learn following facts:
\begin{itemize}
\item For $m_{\tilde{\nu}_1}$ ranging from about $127$ to $133~{\rm GeV}$, the DM mainly co-annihilated with the higgsino-dominated $\tilde{\chi}_1^0$ to achieve the measured density.  In this case, $\sigma_{\tilde{\nu}_1-p}$ can be suppressed to $10^{-51}~{\rm cm^2}$. In contrast, the cross-section is usually larger than $10^{-50}~{\rm cm^2}$ in the Type-I NMSSM for the same mass range. Moreover, the XENON-1T experiment limits $\lambda_\nu \lesssim 0.44$, which is significantly weaker than the limitation on  $\lambda_\nu$ in the Type-I NMSSM\footnote{Note that the Yukawa coupling $\lambda_\nu$ in the Type-I NMSSM corresponds to $2 \lambda_\nu$ in the ISS-NMSSM, which can be inferred by comparing the strength of the quartic scalar coupling $\tilde{\nu} \tilde{\nu} s s$ in the two theories.}.
\item Similar to the Type-I NMSSM, $1 \sigma$ CIs is allowed to be in the mass range $110~{\rm GeV} \lesssim m_{\tilde{\nu}_1} \lesssim 125~{\rm GeV}$, where the DM obtains its density mainly through the process $\tilde{\nu}_1 \tilde{\nu}_1 \to h_1 h_1, h_1 h_2$. The difference is that the CI of the ISS-NMSSM is much broader than that of the Type-I NMSSM, and the cross-section can be significantly lower than the prediction of the Type-I NMSSM.
\item In the case of $m_{\tilde{\nu}_1}\sim 95~{\rm GeV}$, $\tilde{\nu}_1$ annihilated mainly through the channel $\tilde{\nu}_1 \tilde{\nu}_1 \to h_1 h_1$. Correspondingly, the DM density requires $\lambda_\nu \sim 0.35$ (see the formula for the relic density in~\cite{Chang:2013oia,Berlin:2014tja}). This phenomenon is absent in the Type-I NMSSM.
\end{itemize}
The differences reflect the fact that the ISS-NMSSM has greater degrees of freedom for adjusting its parameters to be consistent with relevant experimental constraints than the Type-I NMSSM, and consequently, its DM physics are more flexible. In addition, the left-handed slepton soft-breaking mass $m_{\tilde{l}}$ also plays a role in determining the DM observables through the matrix element $V_{11}$, which is different from the situation for the Type-I NMSSM. In presenting the contours in Fig.~\ref{fig2}, we fixed $m_{\tilde{l}} = 2~{\rm TeV}$.  We verified that the conclusions were not affected by this specific choice. Instead, adopting a smaller $m_{\tilde{l}}$ might slightly improve the fit of the ISS-NMSSM to the DM observables. For example, in the region with $m_{\tilde{\nu}_1} \lesssim 125~{\rm GeV}$, we obtained $\chi^2_{\rm min} \simeq 0.45 $ for the best point with $m_{\tilde{l}} = 2~{\rm TeV}$ and $\chi^2_{\rm min} \simeq 0.02 $ for $m_{\tilde{l}} = 400~{\rm GeV}$.

In Table \ref{table3}, we list the detailed information of the best point in Fig.~\ref{fig2}  (labelled as Point I with corresponding $\chi^2 \simeq 0$)  and a point with $m_{\tilde{\nu}_1} \simeq 96~{\rm GeV}$ and $\chi^2 \simeq 1.4$ (labelled as Point II). From this table, the followings are evident:
\begin{itemize}
\item For Point I, the DM particle is dominated by the $\tilde{x}$ field in its component, while for Point II, it is an equal mix of the $\tilde{\nu}_R$ and $\tilde{x}$ fields\footnote{This fact reflects the general conclusion that the maximal mixing in the sneutrino sector is helpful for the ISS-NMSSM to evade the experimental constraints. We infer this through intensive scans over the parameter space of the model.}.
    In contrast, the DM in the Type-I NMSSM is mainly composed of the $\tilde{\nu}_R$ field.
\item The values $\Delta_{\Omega} = 25$ and  $\Delta_\sigma = 0.08$ for Point I reflect the insensitivity of these quantities to the input parameters. The implication is that there is a large parameter space around the point that satisfies the experimental constraints. We checked that the mass splitting of $\tilde{\nu}_1$ and $\tilde{\chi}_1^0$ was mainly determined $\Delta_{\Omega}$, which was similar to the best points of the Type-I NMSSM.
\item Given $\Delta_{\Omega} = 167$ and  $\Delta_\sigma = 99$, Point II requires considerable tuning to coincide with the experimental results, and thus, it is difficult to obtain in the scan. Our results indicate that, as far as the point is concerned, $\Omega_{th}$ is most sensitive to $\lambda_\nu$ and $A_{\lambda_\nu}$, while $\sigma_{\tilde{\nu}_1-p}$ is most sensitive to $\lambda_\nu$ and $Y_\nu$.
\end{itemize}

\subsection{Effective natural NMSSM scenario}

In the seesaw extensions of the NMSSM, the sparticle's signals may be distinct from those in the NMSSM, and so is the strategy to look for them at the LHC. This feature is relfected in two aspects. One is that, since the sneutrino DM carries a lepton number and has very weak interactions with particles other than the singlet-dominated Higgs boson and massive neutrinos, the decay chain of the sparticles is usually long. Moreover, the decay branching ratio depends not only on the particle mass spectrum but also on new Higgs couplings, such as $Y_\nu$ and $\lambda_\nu$. As a result, the phenomenology of the sparticles is quite complicated. The other aspect arises because both the DM search experiments and the collider experiments relax significantly their constraints on the extensions~\cite{Cao:2019ofo}. Consequently, broad parameter spaces in the NMSSM, which have been excluded by the experiments, are resurrected as experimentally allowed. In particular, the higgsino mass maybe around $100~{\rm GeV}$ to predict the $Z$-boson mass naturally~\cite{Cao:2017cjf,Cao:2018iyk}. This fact makes the phenomenology of the extensions quite rich. Despite these differences, we will show in the following that the phenomenology of the extensions may still mimic that of the NMSSM in a particular case, which we dub the effective natural NMSSM scenario (ENNS).

The ENNS contains only the fields of the NMSSM, with its parameter space automatically satisfying the constraints from DM physics and its potentially significant collider signals self-contained in the framework. The following observations motivate this scenario:
\begin{itemize}
\item From the discussion in previous sections and our Bayesian analysis of the Type-I NMSSM in~\cite{Cao:2018iyk}, the co-annihilation of the sneutrino DM with the lightest neutralino is the most important mechanism to obtain the measured DM density\footnote{Very recently, we performed an analysis of the ISS-NMSSM that was siimlar to our analysis of the Type-I NMSSM. After studying the posterior PDFs of the samples, we determined that this conclusion also applies to the ISS-NMSSM.}.
\item The co-annihilation is insensitive to the parameter $\lambda_\nu$, and it is consistent with the constraints from the DM DD and ID experiments given that $\lambda_\nu$ is not too large.
\item The co-annihilation has the distinct kinematic feature $m_{\tilde{\nu}_1} \simeq m_{\tilde{\chi}_1^0}$, which implies that the DM mass is roughly determined by $m_{\tilde{\chi}_1^0}$ in the NMSSM. In this case, $\tilde{\chi}_1^0$ corresponds to missing momentum at the LHC.
\item As indicated by the best points in previous discussions, $\lambda_\nu$ and $Y_\nu$ are preferred to be less than 0.1. In this case, the $\tilde{\nu}_R$- or $\tilde{x}$-dominated sneutrino couples very weakly with the other particles. For such a sneutrino DM, it intervenes in the phenomenology of the theories mainly by appearing in the decay chain of heavy sparticles.
\end{itemize}

The key features of the ENNS are as follows
\begin{itemize}
\item The lightest neutralino is higgsino-dominated, so the sneutrino DM can co-annihilate with it to obtain the right density. In this case, the higgsino-dominated particles $\tilde{\chi}_{1,2}^0$ and $\tilde{\chi}_1^\pm$ act as the lightest supersymmetric particles of the scenario and are shown as missing momentum at the LHC if the splitting of their masses with the sneutrino DM mass is less than several tens of GeV. This situation is the same as the natural MSSM~\cite{Baer:2012uy}.

    The decay of the singlino-dominated neutralino is somewhat complicated. If its higgsino component is small, it may decay dominantly into DM and a massive neutrino, which will subsequently decay into $W \tau$, $Z \nu$, and $h \nu$ states. This case, however, is of less theoretical interest since this neutralino couples weakly with the other heavier sparticles and has little effect on their decay. Alternatively, if the neutralino contains sizable higgsino components, it prefers to decay into the higgsino-dominated neutralino/chargino plus a vector boson or a Higgs boson.

\item The decays of the gaugino-dominated particles and the colored sparticles are the same as the prediction of the NMSSM with the lightest neutralino being a DM candidate because they do not interact directly with the sneutrinos.

\item The decay modes of charged sleptons are scarcely changed. This can be understood as follows. The extensions predict singlet-dominated particles $\nu_h$ (massive neutrinos) and $\tilde{\nu}$ (sneutrinos), so the left-handed slepton has additional decay channels $\tilde{l}_L^\pm \to \tilde{\nu} H^\pm, \tilde{\nu} W^\pm, \tilde{\chi}_1^\pm \nu_h$. Since $H^\pm$ is preferred to be heavy by the LHC search for charged Higgs bosons and also by $B$-physics measurements, the decay $\tilde{l}_L^\pm \to \tilde{\nu} H^\pm$ is usually kinematically forbidden for a moderately light $\tilde{l}_L$.  For the process $\tilde{l}_L^\pm \to \tilde{\nu} W^\pm$, it proceeds by the small $\tilde{\nu}_L$ component in $\tilde{\nu}$, so its width is suppressed. The channel $\tilde{l}_L^\pm \to \tilde{\chi}_1^\pm \nu_h$ is induced by the Yukawa coupling $Y_\nu$ and can be negligible if $|Y_\nu|$ is much smaller than the magnitude of the gaugino component in $\tilde{\chi}_1^0$, which enables the decay $\tilde{l}_L^\pm \to l^\pm \tilde{\chi}_1^0$ to be dominant~\cite{LDM-27}.  The right-handed sleptons may also decay into these final states. Compared with the left-handed sleptons, the decays must be proceeded by an additional chiral flipping $\tilde{l}_R \to \tilde{l}_L$, so their widths are further suppressed.

    One may also discuss the decay of the left-handed sneutrino, and the conclusion is that its decay pattern changes little for a small $Y_\nu$.

\item The singlet dominated Higgs bosons may be light, which is one of the interesting features in the NMSSM. In this case, one can adjust the parameter $\lambda_\nu$ to enhance $m_{\nu_h}$ so that the decay of the Higgs into $\bar{\nu}_h \nu_h$ is kinematically forbidden. In this case, the decay of the Higgs boson is the same as the NMSSM predictions.

\end{itemize}

We emphasize that the ENNS can be realized by only requiring $2 \kappa \gtrsim \lambda$, and it corresponds to the case of the extensions that is most favored by DM experiments. This situation encourages the study of the phenomenology of the NMSSM without considering the constraints from DM experiments.

\section{Conclusions} \label{Section-conclusion}

With the rapid progress in the DM DD experiments, the sensitivity to DM-nucleon scattering has reached unprecedented precision, which is at the order of $10^{-47}~{\rm cm^2}$ for the SI cross-section and $10^{-42}~{\rm cm^2}$ for the SD cross-section.  These experiments, together with the fruitful LHC experiments, have tightly limited the light higgsino scenario of popular supersymmetric theories such as the MSSM and NMSSM, which customarily take the lightest neutrino as the DM candidate, and thus deteriorate their naturalness in predicting electroweak symmetry breaking. Therefore, it is necessary to seek new mechanisms to suppress the scattering naturally. It is notable that such a mechanism usually alters the DM properties and consequently changes the phenomenology of the traditional theories.
This property may be helpful for the underlying theory to survive the LHC experiments.

After analyzing the fundamental origin of the tight constraints on the MSSM and NMSSM, we realized that, if the DM corresponds to a singlet field under the SM gauge group or at least its singlet component is naturally highly dominant over the other components, the DM-nucleon scattering can be spontaneously suppressed. Based on such observations and our previous work, we pointed out that, in the NMSSM augmented with the Type-I seesaw mechanism or the inverse seesaw mechanism, the lightest sneutrino as a DM candidate can automatically possess this property. Furthermore, due to the simplicity of the theoretical structure, these extensions are economical supersymmetric  theories that naturally suppress the scattering. In addition, we showed that the singlet-dominated Higgs boson in this framework plays a vital role in multiple ways: besides generating the higgsino and massive neutrino masses as well as significantly affecting the SM-like Higgs boson mass, it also mediates the annihilation of the DM or acts as the final state of the annihilation. Consequently, the sneutrino in the extensions is a viable WIMP DM.

Although it is known that the scattering is usually suppressed in the framework, we considered a particular configuration in the Higgs sector that was able to enhance the scattering cross-section to improve our understanding of the DM-nucleon scattering. Our study of this dangerous case revealed that the sneutrino DM could co-annihilate with the lightest neutralino to obtain the measured density. In this case, the constraint from the DD experiment is readily satisfied, and the induced tunings in the DM physics are not serious. Our study of the dangerous case also revealed that the DD experiments were able to exclude the annihilation of the sneutrino DM into a pair of singlet-dominated Higgs bosons as the dominant channel for the DM relic density in the Type-I NMSSM. However, they still allowed the process to account for the density in the ISS-NMSSM. This conclusion shows that the ISS-NMSSM is more flexible in DM physics than the Type-I NMSSM, since the ISS-NMSSM has more theoretical parameters and a more complex structure.

We also discussed the phenomenology of the seesaw extensions of the NMSSM and showed that it might be quite similar to that of the NMSSM in the co-annihilation case, which readily satisfies the constraints from DM experiments. This fact implies that one may ignore the constraints of DM physics on the NMSSM
when studying its phenomenology.

In summary, the WIMP DM (such as the sneutrino discussed herein) in supersymmetric theories is still a good DM candidate that can be naturally consistent with the DM DD and ID experimental results and the LHC search for sparticles even when the higgsino mass is around $100~{\rm GeV}$. Consequently, relevant supersymmetric theories deserve an intensive study. We comment that, although we assume that the sneutrino DM accounts for the total DM density, the conclusion that the seesaw extensions can naturally suppress the DM-nucleon scattering is valid in the multi-component DM case. In this case, this kind of theory is valuable if one of the DM components corresponds to a WIMP, and its scattering with nucleons is found to be tiny.

\section*{Acknowledgements}
This work was supported by the National Natural Science Foundation of China (NNSFC) under Grant No. 11575053.

\begin{figure*}[t]
		\centering
        \resizebox{1.0\textwidth}{!}{
        \includegraphics[width=1.0\textwidth]{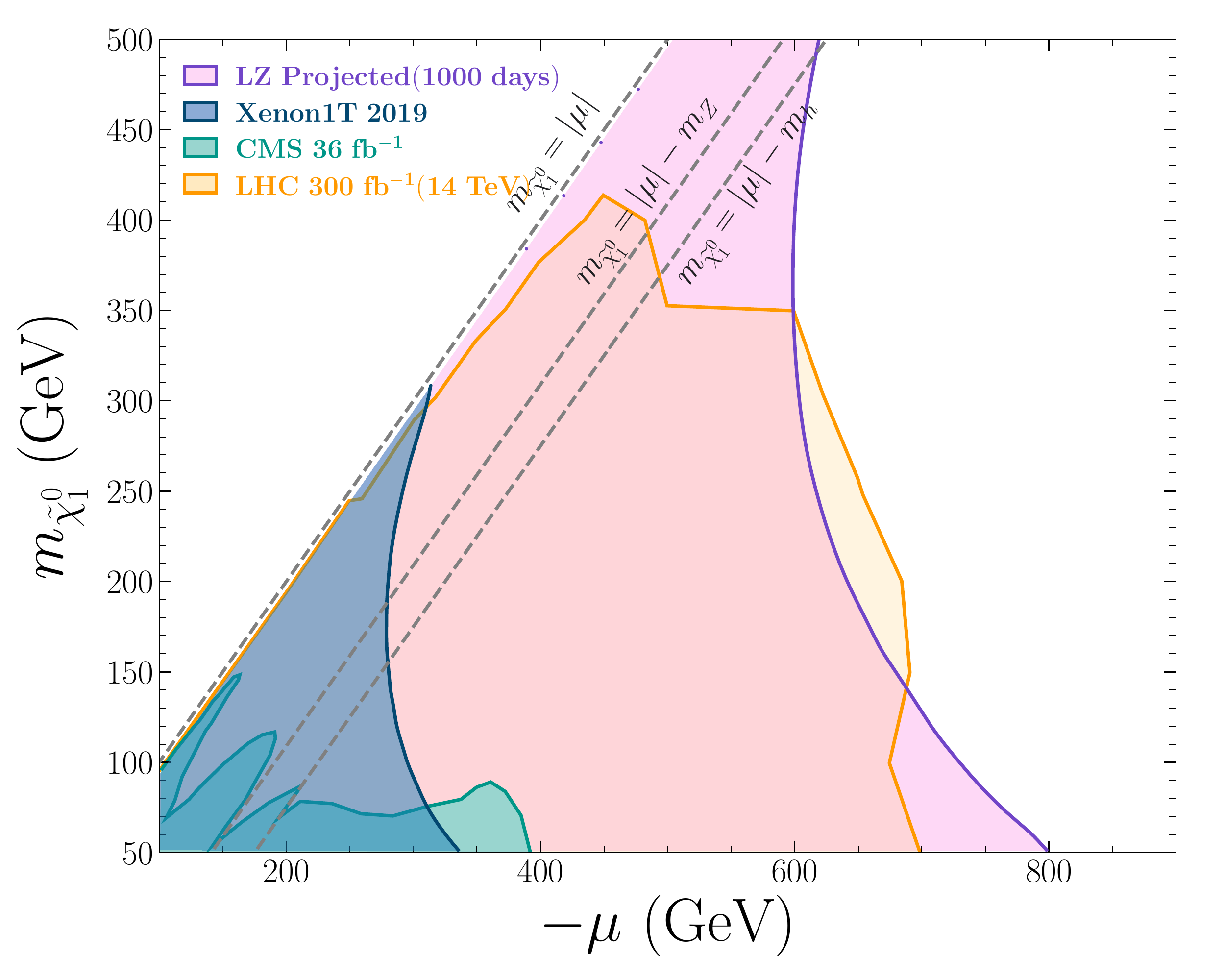}
        \includegraphics[width=1.0\textwidth]{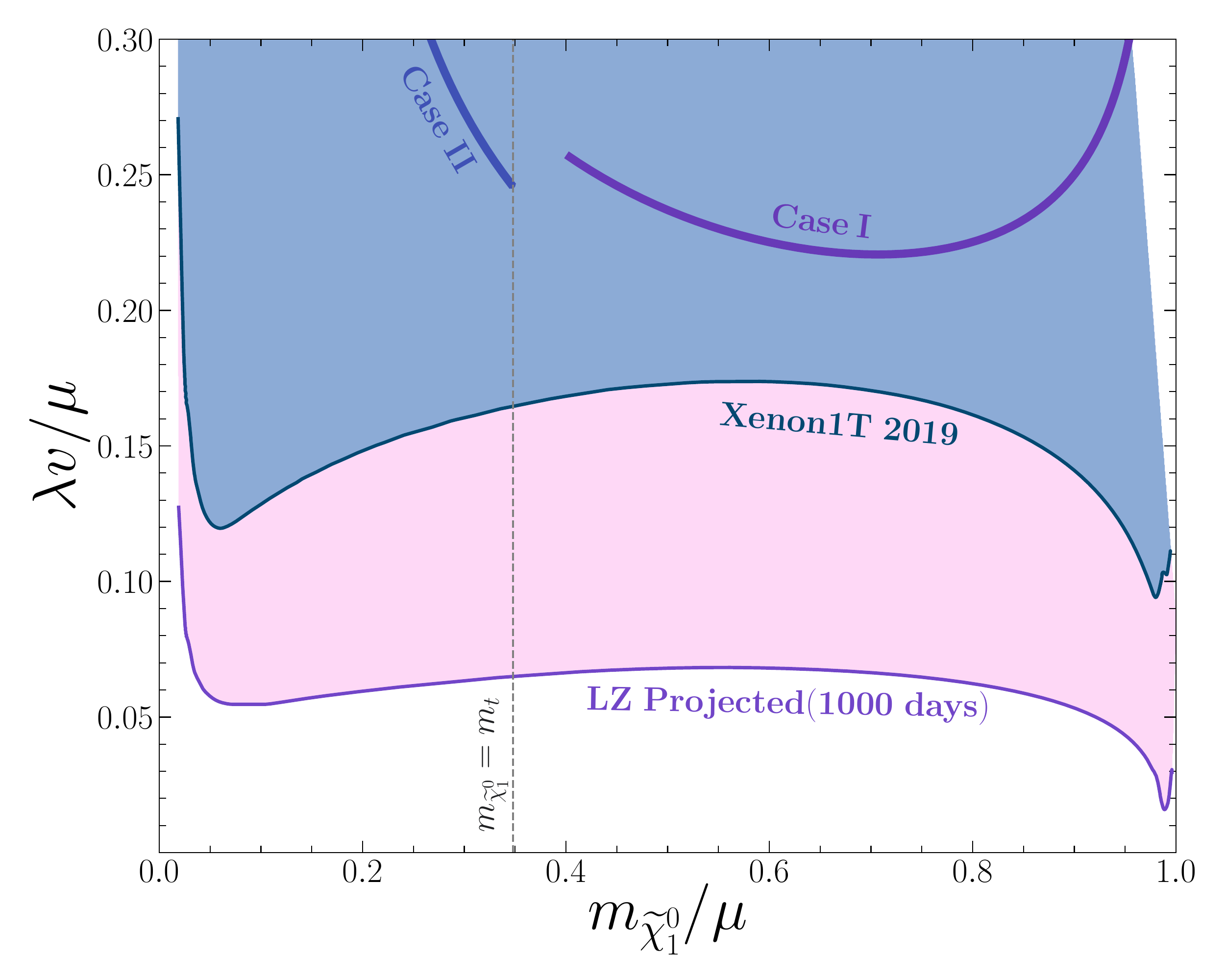}
        }
       \vspace{-0.6cm}
        \caption{Constraints from the detection of the SD DM-nucleon scattering on the blind spot scenarios of the MSSM (left panel) and the NMSSM (right panel). The dark blue region has been excluded by the XENON-1T (2019) experiment, and the purple region can be excluded by the future LZ experiment~\cite{Akerib:2018dfk}. In the left panel, constraints from the CMS search for electroweakinos with $36~{\rm fb^{-1}}$ data at the $13~{\rm TeV}$ LHC as well as the estimated exclusion capabilities with $300~{\rm fb^{-1}}$ data at the $14~{\rm TeV}$ LHC in the same search are also shown, which correspond to the green region and the orange region, respectively. In the right panel, parameter points on the curve labeled by Case I predict the right DM relic density through the annihilation channel $\tilde{\chi}_1^0 \tilde{\chi}_1^0 \to t \bar{t}$, while those on the curve labeled by Case II obtain the correct density through the process $\tilde{\chi}_1^0 \tilde{\chi}_1^0 \to h_s A_s$.     \label{fig3} }
\end{figure*}	

\section{Appendix}

In this appendix, we mainly discuss the constraints of the SD cross-section measurement on the blind spots of the MSSM and the NMSSM. So far, the strictest limit in this regard arises from the analysis of the XENON-1T experiment on DM-neutron scattering~\cite{Aprile:2019dbj}, and the relevant cross-section in the theories is given by~\cite{Badziak:2015exr,Badziak:2017uto}
\begin{eqnarray}
\sigma_{\tilde{\chi}_1^0-n}^{\rm SD} = 3.1 \times 10^{-40}~{\rm cm^2} \times \left (\frac{|N_{13}|^2-|N_{14}|^2}{0.1} \right )^2,
\end{eqnarray}
where $N_{13}$ and $N_{14}$ are the elements of the rotation to diagonalize the neutralino mass matrix, which represent the fractions of the higgsino component in $\tilde{\chi}_1^0$. In the case of $|M_1/\mu | \ll 1$ ($|\lambda v/\mu | \ll 1$) for the MSSM (NMSSM), one may expand $N_{13}$ and $N_{14}$ in powers of  $M_1/\mu$ ($\lambda v/\mu$) to obtain the approximation in Eq.~\ref{MSSM-Z-coupling} (Eq.~\ref{NMSSM-Z-Coupling}). Since the expansion is conditional~\cite{Pierce:2013rda,LDM-27,Cheung:2014lqa} and cannot be applied to all the cases that arise,  we obtain the values of the elements by numerically diagonalizing the neutralino mass matrix in this appendix.

In the theories under consideration, the SI scattering is induced by the SM-like Higgs boson and the other CP-even Higgs bosons. If the latter contribution to the cross-section is negligible, the blind spot condition of the MSSM with a bino-dominated DM can be simplified as $\sin 2 \beta = - m_{\tilde{\chi}_1^0}/\mu$~\cite{Huang:2014xua}. In the heavy wino limit, $ m_{\tilde{\chi}_1^0} = M_1$ by the formula for $m_{\tilde{\chi}_1^0}$ in~\cite{Cheung:2014lqa}. Likewise, the blind spot of the NMSSM with a singlino-dominated DM is characterized by $\sin 2 \beta = m_{\tilde{\chi}_1^0}/\mu$ and $m_{\tilde{\chi}_1^0} = 2 \kappa \mu/\lambda $ in the heavy gaugino limit~\cite{Badziak:2015exr,Cheung:2014lqa}.
We will use these formulas in our discussion.

\subsection{Constraints on blind spot of MSSM}

Given $\sin 2 \beta = - M_1/\mu$ at the blind spot of the MSSM, the neutralino mass matrix depends only on $\mu$ and the gaugino masses $M_{1,2}$. In the left panel of Fig.~\ref{fig3}, we plot the SD constraint of the XENON-1T experiment~\cite{Aprile:2019dbj} in the $m_{\tilde{\chi}_1^0}-\mu$ plane of the MSSM (the dark blue region) as well as the constraint of the future LZ experiment~\cite{Cushman:2013zza,Akerib:2018dfk} in the plane (the purple  region) by setting $M_2 = 5~{\rm TeV}$. These results show that the current bound of the SD cross-section requires $|\mu|$ at the blind spot to be larger than about $300~{\rm GeV}$ regardless of the DM mass, and the future experiments can further exclude $|\mu|$ up to $800~{\rm GeV}$.

As a useful complement to the DD experiments, we also show collider constraints at the $95\%$ confidence level on the blind spot of the MSSM, which arise from the CMS search for electroweakinos through multi-lepton signals of the process $p p \to \tilde{\chi}_{2,3}^0 \tilde{\chi}_1^\pm$ at the $13~{\rm TeV}$ LHC with $36~{\rm fb^{-1}}$ data~\cite{Sirunyan:2018ubx}. We mark the excluded region with green color in the left panel of Fig.~\ref{fig3}. In obtaining this region, we performed detailed simulations of the electroweakino production process in the same way as previously reported~\cite{Cao:2018rix}. We compared our results with the upper bounds of the cross-section for the electroweakino production, which were provided by the CMS collaboration, and found that they roughly coincided. Exclusion capabilities with $300~{\rm fb^{-1}}$ data at the $14~{\rm TeV}$ LHC in the same search were estimated in Fig.~16 of~\cite{Han:2016qtc}, and they are plotted as the orange region in the left panel of Fig.~\ref{fig3}. These results show that $|\mu|$ has been excluded by the LHC experiment up to about $390~{\rm GeV}$ for $m_{\tilde{\chi}_1^0} = 0$ and will be further excluded up to $700~{\rm GeV}$ in the future. We emphasize that, in both calculations of the exclusion capability, the branching ratios ${\rm Br}(\tilde{\chi}_{2,3}^0 \to \tilde{\chi}_{1}^0 Z)$ and ${\rm Br}(\tilde{\chi}_{2,3}^0 \to \tilde{\chi}_{1}^0 h)$ were computed in the decoupling limit $m_A \gg v$, and their effects on the signal event number were taken into account in the simulations.

\subsection{Constraints of SD cross-section on blind spots of NMSSM }

We discuss three cases of the natural NMSSM where the channels $\tilde{\chi}_1^0 \tilde{\chi}_1^0 \to t \bar{t}, h_s A_s, h A_s$ are responsible, respectively, for the measured DM relic density. The common feature of the cases is that the density requires a moderately large $|\lambda v/\mu|$ through an explicit or indirect dependence on the combination,
and under the circumstance that the $h_s$ induced contribution to $\sigma^{\rm SI}_{\tilde{\chi}_1^0-N}$ is negligible, rather strong cancellation between the two terms on the right side of Eq.~(\ref{NMSSM-Higgs-Coupling}) is necessary to render the case consistent with the bound of the SI cross-section (discussed below).  This fact is the basic reason that we considered the blind spot of the SI cross-section and studied the SD constraint for the cases when discussing their compatibility with the DM DD experiments.

The neutralino sector of the NMSSM involves the parameters $\lambda$, $\kappa$, $\tan \beta$, $\mu$, and the gaugino masses $M_{1,2}$~\cite{Ellwanger:2009dp}.  After taking
$M_{1,2} = 5~{\rm TeV}$ and $\tan \beta = m_{\tilde{\chi}_1^0}/\mu = 2 \kappa/\lambda$, the SD cross-section depends only on $\lambda$, $\kappa$, and $\mu$. Considering that the approximation of the $\tilde{\chi}_1^0 \tilde{\chi}_1^0 Z$ coupling in Eq.~(\ref{NMSSM-Z-Coupling}) relies on the combinations $\lambda v/\mu$ and $m_{\tilde{\chi}_1^0}/\mu$
and that a large $\mu$ can relax the constraint, we project the SD constraint of the XENON-1T experiment~\cite{Aprile:2019dbj} onto the $\lambda v/\mu - m_{\tilde{\chi}_1^0}/\mu$ plane by setting $\mu = 500 {\rm GeV}$, which is shown on the right panel of Fig.~\ref{fig3}. This panel indicates that the region of $\lambda v/\mu \gtrsim 0.15$, which corresponds to $\lambda \gtrsim 0.45$ and is marked with dark blue color, has been excluded. We also show the exclusion capability of the future LZ experiment in the same plane, which corresponds to the purple region of the figure. The results indicate that the region of $\lambda v/\mu \gtrsim 0.06$ (corresponding to $\lambda \gtrsim 0.18 $) will be excluded.
Alternatively, if we take a lower value of $\mu$, e.g., $\mu = 300~{\rm GeV}$, the region of $\lambda v/\mu \gtrsim 0.14$ (corresponding to $\lambda \gtrsim 0.23 $) has been excluded, and a broader region bounded by $\lambda v/\mu \gtrsim 0.05$ (corresponding to $\lambda \gtrsim 0.09 $) will be excluded.

Next, we consider the three cases mentioned above.
\begin{itemize}
\item Case I: the annihilation  $\tilde{\chi}_1^0 \tilde{\chi}_1^0 \to t \bar{t}$ is responsible for the DM density.

This case usually occurs when $m_{\tilde{\chi}_1^0} > m_t $~\cite{Badziak:2017uto,Baum:2017enm}. Barring the $s$-channel exchange of a resonant Higgs boson, the annihilation is dominated by the $Z$-mediated contribution, and the density requires the coupling of the DM pair with Goldstone boson to satisfy~\cite{Baum:2017enm}
\begin{eqnarray}
|C_{\tilde{\chi}_1^0 \tilde{\chi}_1^0 G^0} | = \frac{\sqrt{2} m_{\tilde{\chi}_1^0} v}{\mu^2} \lambda^2 \cos 2 \beta \simeq 0.1.  \label{Case1}
\end{eqnarray}
With the blind spot condition, the equation correlates $\lambda$ and $\kappa$ by
\begin{eqnarray}
\kappa^2 \simeq \frac{\lambda^2 \pm \sqrt{\lambda^2 - 0.16}}{8}.
\end{eqnarray}
In the right panel of Fig.~\ref{fig3}, we project this correlation onto the $\lambda v/\mu - m_{\tilde{\chi}_1^0}/\mu$ plane to obtain the curve labeled by Case I. This curve shows that, although the parameter points on it are consistent with the SI constraints, they do not satisfy the density and the SD constraints simultaneously.
This conclusion has been pointed out explicitly in~\cite{Badziak:2017uto}.

As a supplement to the discussion above, we comment on the SI cross-section without the blind spot condition. Eq.~(\ref{Case1}) implies that the first term on the right side of Eq.~(\ref{NMSSM-Higgs-Coupling}) should be larger than $0.1$, while the XENON-1T bound on the SI cross-section requires $|C_{\tilde{\chi}_1^0 \tilde{\chi}_1^0 h}| \lesssim 0.02$ for $m_{\tilde{\chi}_1^0} = 300~{\rm GeV}$. Thus, strong cancellation in Eq.~(\ref{NMSSM-Higgs-Coupling}) is necessary and that is why we considered the blind spot for Case I.

\item Case II: the annihilation  $\tilde{\chi}_1^0 \tilde{\chi}_1^0 \to h_s A_s$ is responsible for the DM density.

Without possible resonant contributions, this annihilation proceeds mainly through the singlino-dominated $\tilde{\chi}_1^0$ in the $t/u$ channel, and it is usually the dominant channel when $m_{\tilde{\chi}_1^0} < m_t $~\cite{Baum:2017enm,Cao:2015loa}. Similar to Case I, the relic density requires~\cite{Baum:2017enm}
\begin{eqnarray}
|C_{\tilde{\chi}_1^0 \tilde{\chi}_1^0 h_s} | &=& |C_{\tilde{\chi}_1^0 \tilde{\chi}_1^0 A_s} | = - \sqrt{2} \kappa \left ( 1- \frac{\lambda^2 v^2}{\mu^2} \right ) + \frac{\lambda^3 v^2}{\sqrt{2} \mu^2} \sin 2 \beta  \nonumber \\
&\simeq& 0.2 \times \left ( \frac{m_{\tilde{\chi}_1^0}}{300~{\rm GeV}} \right )^{1/2},  \label{Case2}
\end{eqnarray}
which translates into the correlation
\begin{eqnarray}
\lambda \kappa (1 - 0.24 \lambda^2 ) \simeq 0.067
\end{eqnarray}
at the blind spot for $\mu = 500~{\rm GeV}$. In the $\lambda v/\mu - m_{\tilde{\chi}_1^0}/\mu$ plane of Fig.~\ref{fig3}, we show the correlation with the curve labelled by Case II. Again, the case cannot satisfy all the constraints of DM physics.

In analogy to the study of Case I, we discuss the situation of the SI cross-section. Eq.~(\ref{Case2}) shows that $\kappa \simeq 0.1$ for $m_{\tilde{\chi}_1^0} = 150~{\rm GeV}$ is able to predict the correct density. Given $m_{\tilde{\chi}_1^0} \simeq 2 \kappa \mu /\lambda $ in case of $\lambda v/\mu \ll 1$~\cite{Cheung:2014lqa}, the first term in Eq.~(\ref{NMSSM-Higgs-Coupling}) is approximately $4 \sqrt{2} \kappa^2 v/m_{\tilde{\chi}_1^0} \simeq 0.066$. This fact together with the XENON-1T bound on the SI cross-section (i.e., $|C_{\tilde{\chi}_1^0 \tilde{\chi}_1^0 h}| \lesssim 0.016$ for $m_{\tilde{\chi}_1^0} = 150~{\rm GeV}$) again inspired us to consider the blind spot.

\item Case III: the annihilation  $\tilde{\chi}_1^0 \tilde{\chi}_1^0 \to h A_s$ is responsible for the DM density.

This channel proceeds mainly by the $t/u$-channel exchange of higgsino-dominated neutralinos~\cite{Baum:2017enm,Cao:2015loa}. The relic density for this case
requires~\cite{Baum:2017enm}
\begin{eqnarray}
\lambda^3 \sin 2 \beta \simeq \left (\frac{\mu}{700~{\rm GeV}} \right )^2,   \label{Case3}
\end{eqnarray}
which corresponds to $\lambda^2 \kappa \simeq 0.25$ at the blind spot for $\mu = 500~{\rm GeV}$. This correlation needs an even larger $\lambda$ than Case II, i.e., $\lambda v/\mu \gtrsim 0.4 $ for $m_{\tilde{\chi}_1^0} < m_t $, and is not shown on the panel of Fig.~\ref{fig3}. In addition, regardless of the blind spot condition, Eq.~(\ref{Case3}) implies that the magnitude of the second term in Eq.~(\ref{NMSSM-Higgs-Coupling}) is approximated by $0.25/\lambda$ for $\mu = 500~{\rm GeV}$. Since the SI constraint of the XENON-1T experiment is equivalent to  $|C_{\tilde{\chi}_1^0 \tilde{\chi}_1^0 h}| \lesssim 0.016$ for $m_{\tilde{\chi}_1^0} = 150~{\rm GeV}$, strong cancellation in Eq.~(\ref{NMSSM-Higgs-Coupling}) is necessary for this case.

\end{itemize}

At the end of this section, we emphasize that the conclusions are based on several hypotheses, i.e., the negligible smallness of the non SM-like Higgs contribution to the SI cross-section, one single annihilation channel responsible for the relic density, and $|\mu| \lesssim 500~{\rm GeV}$. For a general case of the NMSSM, these hypotheses are usually violated. In particular, $\mu$ may be chosen at $1~{\rm TeV}$ if one does not consider the fine tuning. In such a situation, these annihilation channels may still be most important to obtain the density~\cite{Badziak:2017uto,Baum:2017enm}. We also emphasize that, although we take $\mu = 500~{\rm GeV}$ as an example to show the tension between the relic density and the SI/SD cross-section, it holds for $\mu < 500~{\rm GeV}$.


\end{document}